\documentclass[fleqn,usenatbib]{mnras}
\usepackage{newtxtext,newtxmath}
\usepackage{float}
\usepackage[caption = false]{subfig}
\usepackage[T1]{fontenc}
\usepackage{ae,aecompl}

\usepackage{graphicx}	% Including figure files
\usepackage{amsmath}	% Advanced maths commands
\usepackage{amssymb}	% Extra maths symbols

\title[Stellar feedback locally regulates galaxy growth ]{A quantitative demonstration that stellar feedback  locally regulates galaxy growth}

\author[J. Zaragoza-Cardiel et al.]{Javier Zaragoza-Cardiel,$^{1,2}$\thanks{E-mail: javier.zaragoza@inaoep.mx}
Jacopo Fritz,$^{3}$                 
Itziar Aretxaga,$^{1}$             
Yalia D. Mayya,$^{1}$ \newauthor              
Daniel Rosa-Gonz\'alez,$^{1}$     
John E. Beckman,$^{4,5,6}$        
Gustavo Bruzual$^{3}$ and %\newauthor            
Stephane Charlot$^{7}$         
\\
$^{1}$Instituto Nacional de Astrof\'isica, \'Optica y Electr\'onica (INAOE), 
 Luis E. Erro 1, Tonantzintla, Puebla, C.P. 72840, Mexico\\
$^{2}$Consejo Nacional de Ciencia y Tecnolog\'ia, Av. Insurgentes Sur 1582, 03940, Ciudad de M\'exico, Mexico\\
$^{3}$Instituto de Radioastronom\'ia y Astrof\'isica, UNAM, Campus Morelia, 58089 Morelia, Mexico\\
$^{4}$Instituto de Astrof\'isica de Canarias, C/ V\'ia L\'actea s/n, 38205 La Laguna, Tenerife, Spain\\
$^{5}$Department of Astrophysics, University of La Laguna, E-38200 La Laguna, Tenerife, Spain\\
$^{6}$CSIC, 28006 Madrid, Spain\\
$^{7}$Sorbonne Universit\'e, CNRS, UMR7095, Institut d'Astrophysique de Paris, F-75014, Paris, France\\
}

\date{Accepted XXX. Received YYY; in original form ZZZ}

\pubyear{2020}

\begin{document}

\label{firstpage}
\pagerange{\pageref{firstpage}--\pageref{lastpage}}
\maketitle
% Abstract of the paper

\begin{abstract}
We have applied stellar population synthesis 
to  500 pc sized regions in a sample of 102 galaxy discs observed with the MUSE spectrograph. 
We derived the star formation history and analyse specifically the ``recent'' ($20\rm{Myr}$) and ``past'' ($570\rm{Myr}$) age bins. 
Using a star formation self-regulator model we can derive local mass-loading factors, $\eta$ for specific regions, and find  that this factor depends on the local stellar mass surface density, 
$\Sigma_*$,  
in agreement with the predictions form hydrodynamical simulations including supernova feedback. We integrate the local $\eta$-$\Sigma_*$ relation using the stellar mass surface density profiles from the Spitzer Survey of Stellar Structure in Galaxies (S4G) to derive global mass-loading factors, $\eta_{\rm{G}}$, as a function of stellar mass, $M_*$. The $\eta_{\rm{G}}$-$M_*$ relation found  is in very good agreement with hydrodynamical cosmological zoom-in galaxy simulations. The method developed here offers a powerful way of testing different implementations of stellar feedback, to check on how realistic are their predictions.
\end{abstract}

% Select between one and six entries from the list of approved keywords.
% Don't make up new ones.
\begin{keywords}
galaxies:  evolution  -- galaxies:  formation --  galaxies:  star  formation --   galaxies: stellar content
\end{keywords}

%%%%%%%%%%%%%%%%%%%%%%%%%%%%%%%%%%%%%%%%%%%%%%%%%%

%%%%%%%%%%%%%%%%% BODY OF PAPER %%%%%%%%%%%%%%%%%%

\section{Introduction}

Understanding the global star formation process in galaxies is of key importance in the comprehension of galaxy formation and evolution. One of the biggest challenges faced by numerical models of galaxy formation derived directly from cosmological models is to explain why the stellar masses of galaxies are consistently lower than those expected from the simulations \citep{2012RAA....12..917S}. This difference has been bridged by invoking internal mechanisms capable of regulating the star formation rate. Two regimes have been generally used: for massive galaxies their nuclear activity is found to be a mechanism which acts in this way \citep{2018Natur.553..307M}. But for low mass galaxies the star formation itself, through feedback, appears to offer a satisfactory mechanism to reduce the star formation rate, making star formation an inefficient process when comparing the stars which are formed with the availability of gas to form them
\citep{2008AJ....136.2846B,2014MNRAS.445..581H,2019Natur.569..519K}. 
Star formation self-regulates by expelling gas, and the amount of gas that flows out of any system is considered to depend on the mass of stars formed.

The models used to explain the consistently low mean star formation rate (SFR) efficiency 
use sub-grid physics parametrised by a mass loading factor, $\eta$, relating the mass outflow rate $\dot{M}_{\rm{out}}$  and the SFR by $\dot{M}_{\rm{out}}=\eta {\rm{SFR}}$  \citep{2010MNRAS.402.1536S,2013MNRAS.436.3031V,2015ARA&A..53...51S,2018MNRAS.480..800H}. 

This factor can be predicted by modelling the feedback process \citep{2013MNRAS.429.1922C,2015MNRAS.454.2691M,2017ApJ...841..101L}, or inferred from observations \citep{2019MNRAS.490.4368S,2019ApJ...886...74M,2019Natur.569..519K,2020MNRAS.493.3081R}. 
However feedback modelling has many uncertainties, and the required observations are scarce and also subject to uncertainty. The present article marks a significant step in making up for the observational deficiencies.

In order to see whether the star formation at different epochs is correlated and to quantify it by estimating the mass-loading factor we apply an empirical method based on stellar population synthesis and the self-regulator model of star formation, which has been presented previously \citep{2019MNRAS.487L..61Z}.

The star formation self-regulator model 
\citep{2010ApJ...718.1001B,2013ApJ...772..119L,2014MNRAS.444.2071D,2014MNRAS.443..168F,2015MNRAS.448.2126A}
assumes mass conservation for a galaxy, 
  which implies that the change per unit time of the gas mass,  
  $\dot{M}_{\rm{gas}}$, equals the 
inflow rate into the galaxy, $\dot{M}_{\rm{in}}$, minus the gas that goes into star formation, SFR,  
 and the gas which flows out of the galaxy, $\dot{M}_{\rm{out}}$: 
 
 \begin{equation}
  \dot{M}_{\rm{gas}}=\dot{M}_{\rm{in}}-\rm{SFR}(1-R+\eta)\rm{,} 
 \end{equation}

\noindent where $R$ is the fraction of the mass which is returned to the interstellar medium from the stellar population.

The spatially resolved star formation self-regulator model applies to segments of a galaxy \citep{2019MNRAS.487L..61Z}, where by segment we mean any spatially resolved region of a galaxy.
In these resolved regions we also assume conservation of mass:   
the time change of the gas mass surface density in a segment,  $\dot{\Sigma}_{\rm{gas}}$,  
is equal to the surface density of the net gas flow rate, $\dot{\Sigma}_{\rm{net\thinspace flow}}$, 
minus the surface density of gas that goes into new stars through star formation, ${\Sigma_{\rm{SFR}}}$, and 
 minus the surface density of gas that is 
expelled from the segment by stellar processes, $\dot{\Sigma}_{\rm{out}}$:

\begin{equation}
  \dot{\Sigma}_{\rm{gas}}={\dot{\Sigma}}_{\rm{net \thinspace flow}}-{\Sigma_{\rm{SFR}}}(1-R+\eta)
   \label{eq_bath_res}
 \end{equation}

\noindent where $R$ is the fraction of the mass that is returned to the interstellar medium, 
and 

\begin{equation}
 {\dot{\Sigma}}_{\rm{out}}=\eta{\Sigma_{\rm{SFR}}}.
 \label{eq_loading_res}
\end{equation}

 This model allows us to relate the star formation rate surface density in a segment, $\Sigma_{\rm{SFR}}$, to the change in gas mass in that segment. 
The complex processes of stellar feedback are parameterized 
by the mass-loading factor:  
$\dot{\Sigma}_{\rm{out}}=\eta {\Sigma_{\rm{SFR}}}$.

We present the galaxy sample and the data in section \S2. In section \S3 we give the stellar population synthesis fits and also fit the observables to the star formation self-regulator model. In section \S4 we show the results obtained,  and the variation of $\eta$, while in section \S5 we convert local values of $\eta$ into global ones. We discuss our results 
in section \S6 and present our conclusions in section \S7.

\section{Galaxy Sample and data}

\subsection{Galaxy sample}
% \begin{landscape}

\begin{table*}
% \begin{sidewaystable*}[h!]
 \centering 
\caption{Galaxy sample.} 
\label{tab_sample} 
\begin{tabular}{rrrrrr}  
\hline 
Galaxy identifier & PGC identifier $^a$ & $D$ $^b$ & z $^c$ & Type $^d$ & i $^e$   \\ & & Mpc & &  & $^{\circ}$  \\ 
\hline 
pgc33816 & PGC33816 & 23.6 & 0.005187 & 7.0 & 19.9 \\ 
eso184-g082 & PGC63387 & 35.2 & 0.00867 & 4.1 & 32.6 \\ 
eso467-062 & PGC68883 & 57.5 & 0.013526 & 8.6 & 49.9 \\ 
ugc272 & PGC1713 & 55.6 & 0.012993 & 6.5 & 70.7 \\ 
ngc5584 & PGC51344 & 23.1 & 0.005464 & 5.9 & 42.4 \\ 
eso319-g015 & PGC34856 & 37.5 & 0.009159 & 8.6 & 54.2 \\ 
ugc11214 & PGC61802 & 38.0 & 0.008903 & 5.9 & 16.5 \\ 
ngc6118 & PGC57924 & 20.5 & 0.005247 & 6.0 & 68.7 \\ 
ic1158 & PGC56723 & 24.5 & 0.006428 & 5.1 & 62.2 \\ 
ngc5468 & PGC50323 & 30.0 & 0.00948 & 6.0 & 21.1 \\ 
eso325-g045 & PGC50052 & 75.9 & 0.017842 & 7.0 & 40.2 \\ 
ngc1954 & PGC17422 & 38.0 & 0.010441 & 4.4 & 61.5 \\ 
ic5332 & PGC71775 & 9.9 & 0.002338 & 6.8 & 18.6 \\ 
ugc04729 & PGC25309 & 57.0 & 0.013009 & 6.0 & 35.2 \\ 
ngc2104 & PGC17822 & 16.4 & 0.003873 & 8.5 & 83.6 \\ 
eso316-g7 & PGC28744 & 47.5 & 0.01166 & 3.3 & 70.0 \\ 
eso298-g28 & PGC8871 & 70.1 & 0.016895 & 3.8 & 64.4 \\ 
mcg-01-57-021 & PGC69448 & 30.6 & 0.009907 & 4.0 & 52.2 \\ 
pgc128348 & PGC128348 & 61.1 & 0.014827 & 5.0 & 36.7 \\ 
pgc1167400 & PGC1167400 & 60.0 & 0.01334 & 4.0 & 30.5 \\ 
ngc2835 & PGC26259 & 10.1 & 0.002955 & 5.0 & 56.2 \\ 
ic2151 & PGC18040 & 30.6 & 0.010377 & 3.9 & 61.5 \\ 
ngc988 & PGC9843 & 17.3 & 0.005037 & 5.9 & 69.1 \\ 
ngc1483 & PGC14022 & 16.8 & 0.003833 & 4.0 & 37.3 \\ 
ngc7421 & PGC70083 & 24.2 & 0.005979 & 3.7 & 36.2 \\ 
fcc290 & PGC13687 & 19.0 & 0.004627 & 2.1 & 48.1 \\ 
ic344 & PGC13568 & 75.6 & 0.018146 & 4.0 & 60.7 \\ 
ngc3389 & PGC32306 & 21.4 & 0.004364 & 5.3 & 66.2 \\ 
eso246-g21 & PGC9544 & 76.6 & 0.018513 & 3.0 & 52.4 \\ 
pgc170248 & PGC170248 & 85.1 & 0.019163 & 4.7 & 76.4 \\ 
ngc7329 & PGC69453 & 45.7 & 0.010847 & 3.6 & 42.7 \\ 
ugc12859 & PGC72995 & 78.3 & 0.018029 & 4.0 & 72.8 \\ 
ugc1395 & PGC7164 & 74.1 & 0.017405 & 3.1 & 55.1 \\ 
ngc5339 & PGC49388 & 27.0 & 0.009126 & 1.3 & 37.5 \\ 
ngc1591 & PGC15276 & 55.8 & 0.013719 & 2.0 & 56.8 \\ 
pgc98793 & PGC98793 & 55.2 & 0.01292 & 5.0 & 0.0 \\ 
ugc5378 & PGC28949 & 56.5 & 0.01388 & 3.1 & 64.1 \\ 
ngc4806 & PGC44116 & 29.0 & 0.008032 & 4.9 & 32.9 \\ 
ngc1087 & PGC10496 & 14.4 & 0.00506 & 5.2 & 54.1 \\ 
ngc4980 & PGC45596 & 16.9 & 0.004767 & 1.1 & 71.5 \\ 
ngc6902 & PGC64632 & 46.6 & 0.009326 & 2.3 & 40.2 \\ 
ugc11001 & PGC60957 & 63.3 & 0.01406 & 8.1 & 78.7 \\ 
ic217 & PGC8673 & 27.0 & 0.006304 & 5.8 & 82.6 \\ 
eso506-g004 & PGC39991 & 57.5 & 0.013416 & 2.6 & 67.2 \\ 
ic2160 & PGC18092 & 64.7 & 0.015809 & 4.6 & 62.7 \\ 
ngc1385 & PGC13368 & 22.7 & 0.005 & 5.9 & 52.3 \\ 
mcg-01-33-034 & PGC43690 & 32.0 & 0.008526 & 2.1 & 56.6 \\ 
\end{tabular} 
\\
\hrulefill
\\
{\raggedright 
$^a$ Principal General Catalog of Galaxies identifier from Hyperleda database \citep{2003A&A...412...45P}.  
$^b$ Distance from the z=0 Multi-wavelength Galaxy Synthesis (z0MGS from \cite{2019ApJS..244...24L}, when available) and HyperLeda database best homogenized distances \citep{2014A&A...570A..13M}.  
$^c$ Redshift, from Nasa Ned.  
$^d$ Numerical morphologycal type, from the HyperLeda database. 
$^e$ Inclination from the HyperLeda database.  }   \\
\hrulefill 
% \end{sidewaystable*}
\end{table*}
% \end{landscape}

\begin{table*}
% \begin{sidewaystable*}[h!]
 \centering 
\contcaption{} 
% \label{tab_sample} 
\begin{tabular}{rrrrrr}  
\hline 
Galaxy identifier & PGC identifier $^a$ & $D$ $^b$ & z $^c$ & Type $^d$ & i $^e$   \\ & & Mpc & &  & $^{\circ}$  \\ 
\hline 
ngc4603 & PGC42510 & 33.1 & 0.008647 & 5.0 & 44.8 \\ 
ngc4535 & PGC41812 & 15.8 & 0.006551 & 5.0 & 23.8 \\ 
ngc1762 & PGC16654 & 76.5 & 0.015854 & 5.1 & 51.5 \\ 
ngc3451 & PGC32754 & 26.1 & 0.00445 & 6.5 & 62.7 \\ 
ngc4790 & PGC43972 & 15.3 & 0.004483 & 4.8 & 58.8 \\ 
ngc3244 & PGC30594 & 42.7 & 0.009211 & 5.6 & 49.3 \\ 
ngc628 & PGC5974 & 9.8 & 0.002192 & 5.2 & 19.8 \\ 
pgc30591 & PGC30591 & 35.5 & 0.006765 & 6.8 & 86.6 \\ 
ngc5643 & PGC51969 & 11.8 & 0.003999 & 5.0 & 29.6 \\ 
ngc1309 & PGC12626 & 24.1 & 0.007125 & 3.9 & 21.2 \\ 
ngc1084 & PGC10464 & 17.3 & 0.004693 & 4.8 & 49.9 \\ 
ngc7580 & PGC70962 & 65.3 & 0.01479 & 3.0 & 36.5 \\ 
ngc692 & PGC6642 & 87.9 & 0.021181 & 4.1 & 45.2 \\ 
eso462-g009 & PGC64537 & 83.2 & 0.019277 & 1.1 & 58.8 \\ 
ic5273 & PGC70184 & 14.7 & 0.004312 & 5.6 & 50.8 \\ 
pgc3140 & PGC3140 & 81.3 & 0.019029 & 1.4 & 62.7 \\ 
ic1553 & PGC1977 & 35.0 & 0.00979 & 7.0 & 78.6 \\ 
ugc11289 & PGC62097 & 59.7 & 0.013333 & 4.5 & 53.7 \\ 
ic4582 & PGC55967 & 37.3 & 0.007155 & 3.8 & 83.1 \\ 
ngc2466 & PGC21714 & 73.1 & 0.017722 & 5.0 & 16.0 \\ 
eso443-21 & PGC44663 & 41.9 & 0.009404 & 5.7 & 79.0 \\ 
ic4452 & PGC51951 & 65.3 & 0.014337 & 1.3 & 20.6 \\ 
eso498-g5 & PGC26671 & 40.7 & 0.008049 & 4.3 & 41.8 \\ 
eso552-g40 & PGC16465 & 95.5 & 0.022649 & 2.1 & 54.4 \\ 
eso163-g11 & PGC21453 & 33.0 & 0.009413 & 3.0 & 70.9 \\ 
ngc7582 & PGC71001 & 18.7 & 0.005254 & 2.1 & 68.0 \\ 
ngc1620 & PGC15638 & 39.6 & 0.011715 & 4.5 & 81.2 \\ 
ic1320 & PGC64685 & 73.6 & 0.016548 & 2.9 & 58.1 \\ 
ngc3393 & PGC32300 & 52.8 & 0.012509 & 1.2 & 30.9 \\ 
ngc2370 & PGC20955 & 79.8 & 0.018346 & 3.4 & 56.8 \\ 
ngc4981 & PGC45574 & 21.0 & 0.005604 & 4.0 & 44.7 \\ 
ngc3783 & PGC36101 & 25.1 & 0.00973 & 1.4 & 26.6 \\ 
ngc1285 & PGC12259 & 74.1 & 0.017475 & 3.4 & 59.3 \\ 
ngc5806 & PGC53578 & 26.2 & 0.004533 & 3.2 & 60.4 \\ 
eso018-g018 & PGC26840 & 71.1 & 0.017572 & 4.2 & 38.9 \\ 
ngc6754 & PGC62871 & 38.4 & 0.010864 & 3.2 & 61.0 \\ 
ic2560 & PGC29993 & 32.5 & 0.009757 & 3.4 & 65.6 \\ 
ngc7140 & PGC67532 & 36.0 & 0.009947 & 3.8 & 49.6 \\ 
ngc3464 & PGC833131 & 52.8 & 0.012462 & 4.9 & 50.8 \\ 
mcg-02-13-38 & PGC16605 & 55.2 & 0.013293 & 1.2 & 73.6 \\ 
ngc1590 & PGC15368 & 55.2 & 0.012999 & 5.0 & 27.9 \\ 
pgc8822 & PGC8822 & 74.1 & 0.017555 & 5.0 & 58.2 \\ 
ngc7721 & PGC72001 & 21.2 & 0.006721 & 4.9 & 81.4 \\ 
pgc28308 & PGC28308 & 43.0 & 0.00907 & 6.7 & 85.5 \\ 
ngc1137 & PGC10942 & 42.8 & 0.010147 & 3.0 & 59.5 \\ 
eso478-g006 & PGC8223 & 74.8 & 0.017786 & 4.2 & 57.7 \\ 
ngc1448 & PGC13727 & 16.8 & 0.003896 & 6.0 & 86.4 \\ 
ngc3278 & PGC31068 & 42.7 & 0.009877 & 5.1 & 41.0 \\ 
ngc4030 & PGC37845 & 19.0 & 0.004887 & 4.0 & 47.0 \\ 
ngc3363 & PGC32089 & 85.1 & 0.019233 & 3.5 & 45.3 \\ 
ngc7780 & PGC72775 & 76.2 & 0.017195 & 2.0 & 61.2 \\ 
ic1438 & PGC68469 & 42.5 & 0.008659 & 1.2 & 23.8 \\ 
ngc4666 & PGC42975 & 15.7 & 0.005101 & 5.0 & 69.6 \\ 
ngc7396 & PGC69889 & 71.8 & 0.016561 & 1.0 & 59.5 \\ 
ngc716 & PGC6982 & 65.9 & 0.015204 & 1.1 & 75.9 \\ 
\hline 
\end{tabular} 
\end{table*}

A significant number of galaxies have been observed with the MUSE instrument on the VLT in different surveys \citep{2017ApJ...844...48P,2018A&A...609A.119S,2019ApJ...887...80K,2019MNRAS.484.5009E,2020AJ....159..167L}. 
To use these observations we build our sample using the Hyperleda database and looking for them in the MUSE archive.
To be able to apply the method for a given galaxy, 
we need to resolve the galaxy at a specific spatial scale. 
Based on results of NGC 628 \citep{2019MNRAS.487L..61Z}, we choose the 500pc scale to study 
the star formation self-regulation so we are limited to galaxies 
closer than 100Mpc to resolve 500pc at $1\thinspace\rm{arcsec}$ resolution. We also need enough  ($\sim16$)
resolution elements, so very nearby galaxies with low number of 500pc resolution elements are not useful. We will divide the MUSE field of view in squares, so we will need at least $4\times4$ 500pc squares per galaxy, limiting us to galaxies further away than 7Mpc. 

We need galaxies with recent star formation to study star formation self-regulation. To ensure that we will detect recent star formation, we just consider Sa or later types morphology (Hubble type $T\geq1.0$ in Hyperleda). We discard edge-on galaxies ($i=90\rm{\rm{\deg}}$), galaxies classified as multiple, Irregulars (Hubble type $T\geq9$ in Hyperleda), and LIRGs (in NASA Ned). We just select galaxies with declination lower than $45\rm{\deg}$N to be observable from Paranal Observatory. 

The SQL (Structured Query Language) search through Hyperleda \footnote{\url{http://leda.univ-lyon1.fr/fullsql.html}} selects 13636 galaxies, of which 164 have been observed with MUSE on the VLT and have publicly available data with an exposure time at least of 1600 seconds. We also removed galaxies in Arp \citep{1966ApJS...14....1A}, Vorontsov-Velyaminov \citep{1959VV....C......0V}, and Hickson Compact Group \citep{1982ApJ...255..382H} catalogs, to get rid out of strong external effects on the star formation history (SFH) and gas flows due to interactions. We have a total of 148 galaxies satisfying these conditions in the public MUSE archive. Of these, 9 galaxies did not pass our requirements in a spectral inspection by eye, because of clear spectral artifacts, or not having enough H$\alpha$ emission in the pointing (MUSE has a square $1\rm{arcmin}\times1\rm{arcmin}$ FOV). We initially analysed the single stellar populations (SSP's)  of the remaining 139 galaxies, to apply the method described in this article. Since the method requires enough regions to be include in the analysis, we set this limit to 16 (4x4).  However each of the 16 regions sampled per galaxy needs sufficient current SFR, sufficient signal to noise, and that can be properly reproduced with stellar population synthesis models. Finally, only 102 galaxies satisfied all the conditions allowing us to estimate $\eta$. We present their parameters in Table \ref{tab_sample}.

\subsection{Muse spectral data}

\begin{figure}

\includegraphics[width=0.42\textwidth]{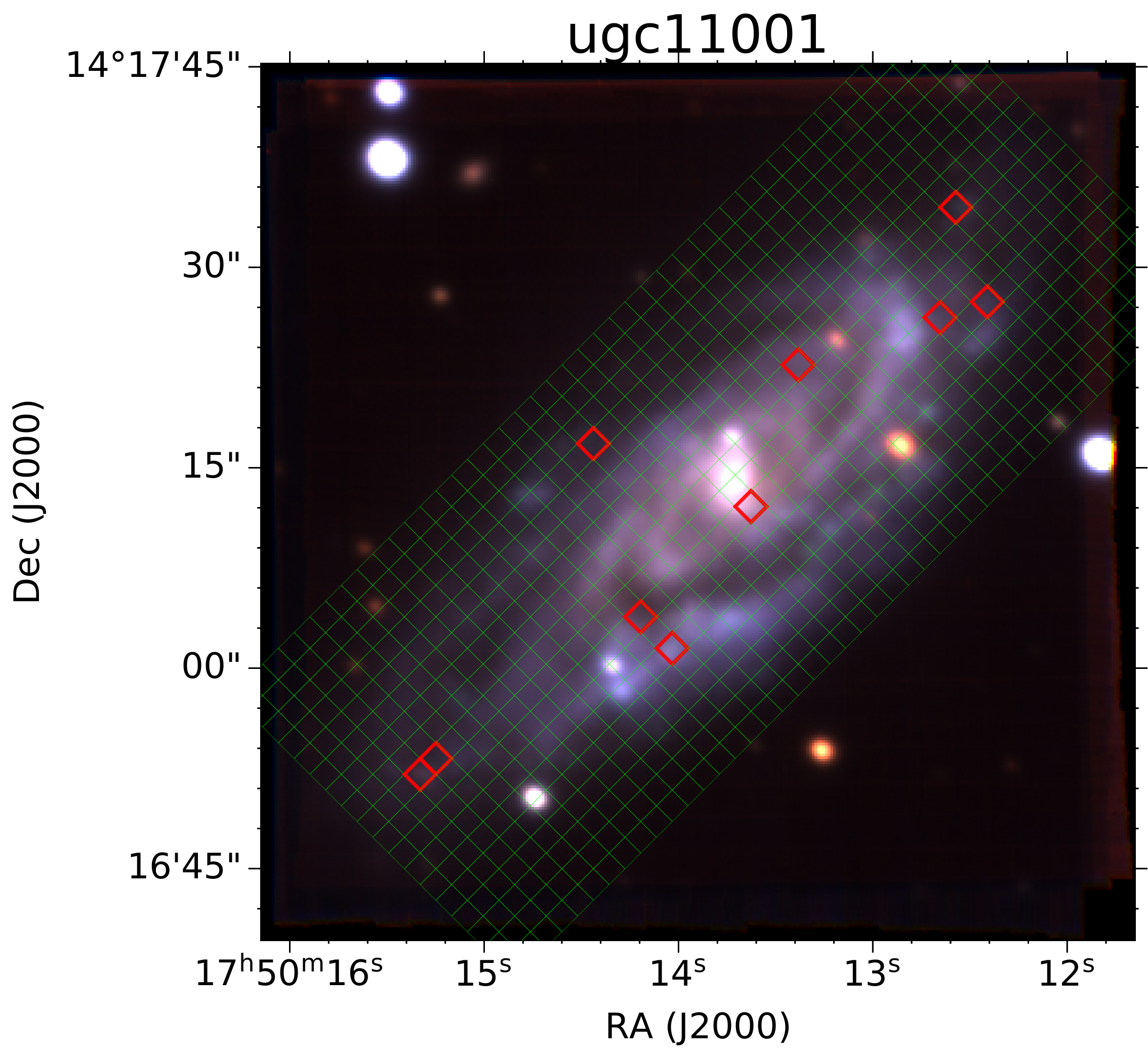}

\caption{Colour composite RGB image recovered from MUSE data of one of the studied galaxies, UGC 11001. The red, green and blue images used to create RGB are obtained by integrating MUSE spectra in R, V and B filters, respectively. 
500pc wide regions where spectra was extracted are overplotted as green 
and the regions identified as those on the envelope (defined above in section \S4) are marked as red squares.}
\label{fig_colormuse}
\end{figure}

We use the MUSE  \citep{2010SPIE.7735E..08B} reduced publicly available data for the galaxies  
listed in Table \ref{tab_sample}, from the ESO 
archive\footnote{\url{http://archive.eso.org/wdb/wdb/adp/phase3_spectral/form?collection_name=MUSE}}. 

We first made a visual inspection to remove galaxies with no H$\alpha$ emission in the MUSE pointing, and thus to select MUSE fields where H$\alpha$ was observed, to be able to estimate recent star formation. After delimiting the regions with recent star formation, we divide each field into an integer number of observing squares, giving us  squares with the closest (and larger than) size value to 500 pc. 
We show an example in Fig. \ref{fig_colormuse}, we do not use the squares outside the MUSE pointing. 
We choose 500 pc because it gives us a scale on which, from previous work, we expect to observe the self-regulation of star formation \citep{2019MNRAS.487L..61Z}, and it allows us to include galaxies at distances of up to 100 Mpc where 500 pc corresponds to 1 arcsec.  
We also need the foreground stars to be masked. We extract the spectrum for each defined region, correct it for Galactic extinction, and associate each with a redshift estimate, using the H$\alpha$ 
or [NII] at $6583.4\rm{\AA}$ if the later has a stronger peak than the former. We next  estimate the [NII]/H$\alpha$ and the [OIII]/H$\beta$ flux ratios, and remove the regions which are classified as Seyfert-LINER in the BPT diagram  \citep{2006MNRAS.372..961K}.  

\section{Stellar population synthesis and model fits}

\subsection{Stellar population synthesis}
\begin{figure}
\begin{center}
\includegraphics[width=0.5\textwidth]{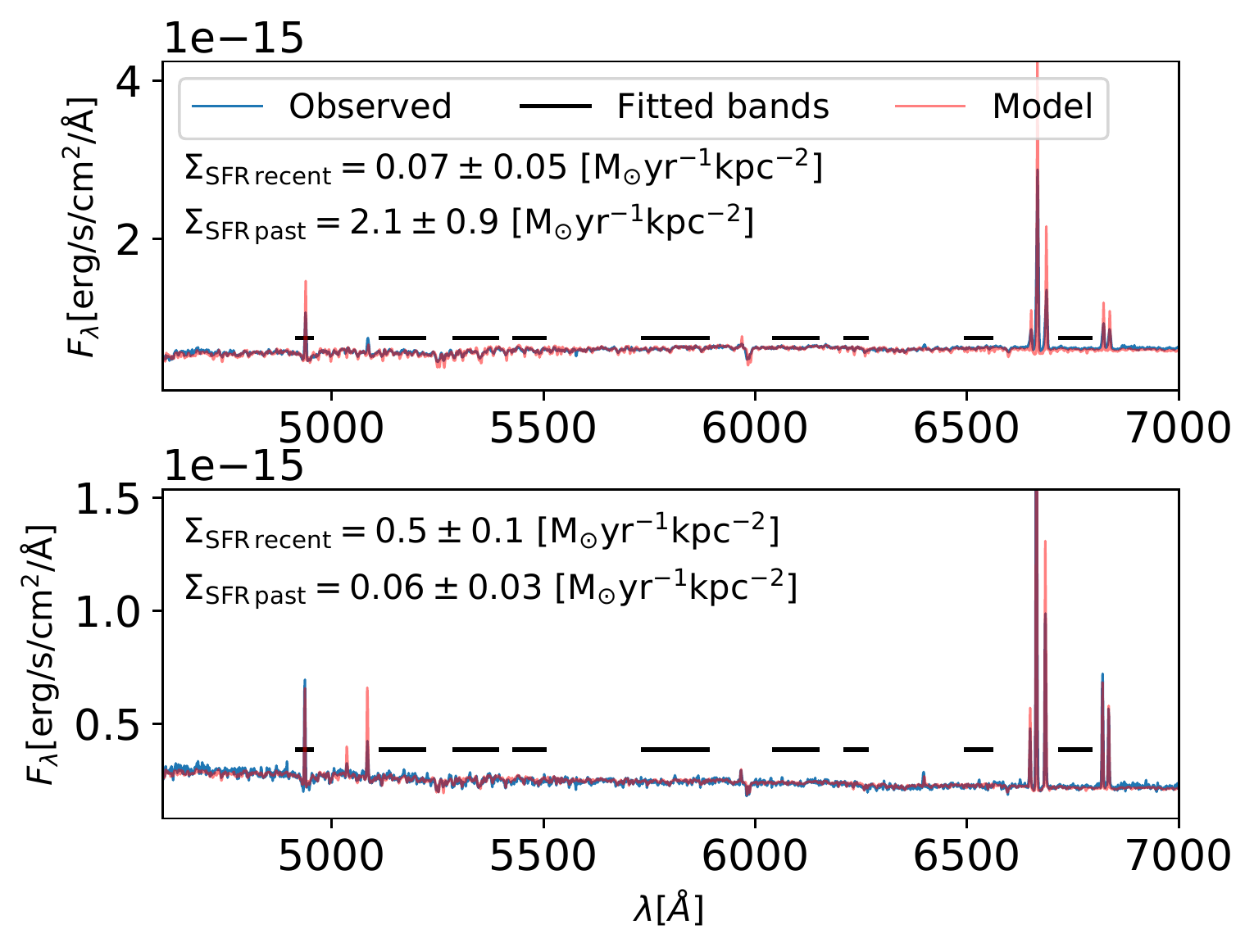} 
\end{center}
\caption{Two characteristic spectra for the galaxy NGC 716. Observed spectra are shown as blue lines. Top: the total star formation is dominated by the past star formation. Bottom: the total star formation is dominated by the recent star formation. The model spectrum which best fits the observed spectrum is shown as a red line. The continuum bands used to fit the observed spectrum to the combination of SSPs are shown as black lines.      
}
\label{fig_specfit}
\end{figure}

We use SINOPSIS code \footnote{\url{https://www.irya.unam.mx/gente/j.fritz/JFhp/SINOPSIS.html}} 
 \citep{2007A&A...470..137F,2017ApJ...848..132F} to fit combinations of SSPs to the 
observed spectra. SINOPSIS fits equivalent widths of emission and absorption lines, as well as 
defined continuum bands. In this work, we use the H$\alpha$ and H$\beta$ equivalent widths, and 
the 9 continuum bands shown in Fig. \ref{fig_specfit}, where we show two observed spectra of the galaxy NGC 716 and the resulted fits as an example.

We use the updated version of the Bruzual \& Charlot models \citep{2019MNRAS.483.2382W}. We used SSPs of 3 metallicities ($Z=0.004$, $Z=0.02$, and  $Z=0.04$) 
in 12 age bins (2 Myr, 4 Myr, 7 Myr, 20 Myr, 57 Myr, 200 Myr, 
570 Myr, 1 Gyr, 3 Gyr, 5.75 Gyr, 10 Gyr, and 14 Gyr). We assume a free form of SFH, the Calzetti dust attenuation law  \citep{2000ApJ...533..682C}, and the Chabrier 2003 IMF  \citep{2003PASP..115..763C} for stellar masses between $0.1\rm{M_{\odot}}$ 
and $100\rm{M_{\odot}}$. The emission lines for the SSPs younger than 20 Myr are computed 
using the photoionisation code 
{\sc{Cloudy}}  \citep{1993hbic.book.....F,1998PASP..110..761F,2013ApJ...767..123F}, 
assuming case B recombination  \citep{1989agna.book.....O}, 
an electron temperature of $10^4 \rm{K}$, an electron density of $100\rm{cm}^{-3}$, 
and a gas cloud with an inner radius of $10^{-2} \rm{pc}$ 
 \citep{2017ApJ...848..132F}.

SINOPSIS uses the degeneracies 
between age, metallicity, and dust attenuation, to compute the uncertainties in the 
derived parameters  \citep{2007A&A...470..137F}.

We rebin the different age bins into 4 bins: at $20\rm{Myr}$, $570\rm{Myr}$, $5.7\rm{Gyr}$, 
and $14\rm{Gyr}$. 
Simulated and observed spectra have been used to prove the validity of using SINOPSIS to recover these 
4 age bins \citep{2007A&A...470..137F,2011A&A...526A..45F}. Additionally, SINOPSIS and similar codes 
have shown the reliability of recovering the SFH using synthesis of SSPs in, at least, 4 age bins \citep{2005MNRAS.358..363C,2007A&A...470..137F,2011A&A...526A..45F,2016RMxAA..52...21S}. However, since we are interested in the recent star formation variations, we  consider only the two most recent age bins, $20\rm{Myr}$, $570\rm{Myr}$, and call them recent, and past age bins, respectively. In this way we recover the recent, and the past star formation rate surface densities, $\Sigma_{\rm{SFR\thinspace recent}}$ and $\Sigma_{\rm{SFR\thinspace past}}$, 
which improves the confidence in the results presented in this work, since the two most recent age bins are better constrained than the oldest ones.

In order to use regions with a meaningful
 result, we take into account only regions with a signal to noise ratio (SNR) larger than 20 over the $[5350-5420]\thinspace \rm{\AA}$ range, and $\chi^2<3$. Due to IMF sampling effects, we also consider only regions where the recent SFR is larger than $10^{-3}\rm{M_{\odot}/yr}$ and the past SFR is larger than $10^{-5}\rm{M_{\odot}/yr}$.

Because we are limited to galaxies from Hubble Type Sa to Sdm, also excluding  interacting galaxies and (U)LIRGs, the galaxy sample, by construction, is defined by galaxies that are probably on the star formation galaxy main sequence, and probably evolved via secular evolution in the studied age range (last 570 Myr), where 
by secular evolution we mean evolution dominated by slow processes (slower than many galaxy rotation periods \citet{2004ARA&A..42..603K}). 
The galaxies probably evolved through more violent episodes in the past, but we are not affected by them in the studied age range.
Nevertheless, individual zones such as the centres of the galaxies, might have evolved via rapid evolution due to high gas flows even in the studied age range. Because of this, we removed regions whose centres are at a distance of 500 pc or less from the centre of the galaxy, as well as regions 
having a very high recent SFR compared to the rest of the galaxy, specifically, we removed regions having $\Sigma_{\rm{SFR\thinspace recent}}$ larger than $\overline{\Sigma}_{\rm{SFR\thinspace recent}}+3\sigma_{\Sigma_{\rm{SFR\thinspace recent}}}$ for each galaxy. We will discuss how affects the results the removal of very high recent SFR regions in the discussion section (\S6.4).

\subsection{Fitting data to self-regulator the model}

We have made the same assumptions made in \citet{2019MNRAS.487L..61Z} in order to fit 
our observables to the self-regulator model. For completeness, we briefly describe them here.

The self-regulator model (Eq. \ref{eq_bath_res}) is valid for a star or a group of co-rotating stars in the galaxy such as a massive star cluster (>500$\rm{M_{\odot}}$ \citet{2003ARA&A..41...57L}). Assuming $\eta$ constant, Eq. \ref{eq_bath_res} is linear, so we can add up regions obeying that equation, and still obey the equation. In this context, the mass-loading factor would be representative of massive star clusters scales ($\sim$pc \citet{2003ARA&A..41...57L}). Although we find below that $\eta$ varies (Eq. \ref{eq_eta_den}), the variation is smooth enough to consider it approximately constant here. 
Therefore, the group of stars which are massive enough to produce bound clusters can be considered as a whole, while the less massive ones are splitted into individual stars. Feedback between different regions is then not considered here. We assume that our 500 pc wide regions are made of individual smaller regions obeying Eq \ref{eq_bath_res}, so we can rewrite Eq. \ref{eq_bath_res} to be valid for our larger regions as the average of individual regions:

\begin{equation}
   \dot{\overline{{\Sigma}}}_{\rm{gas}}=\dot{\overline{{\Sigma}}}_{\rm{net \thinspace flow}}-\overline{\Sigma}_{\rm{SFR}}(1-R+\eta).
   \label{eq_bath_res_sum}
\end{equation}

We already showed in \cite{2019MNRAS.487L..61Z} that the resulted mass-loading factor was independent of the chosen scale (from 87pc to 1kpc) in NGC 628. Hence, regions can be added up while Eq. \ref{eq_bath_res_sum} is still valid.   

The value of the $\overline{\Sigma}_{\rm{SFR\thinspace past}}$ we are able to measure is a time average over 550Myr. Since Eq. \ref{eq_bath_res_sum} is linear, we can substitute the time differentials by time average values over our age bin, and we will not be affected by possible bursts of the star formation, as long as the variation of $\eta$ is small enough (as we do find below).

The net gas flow rate surface density, $\dot{\overline{{\Sigma}}}_{\rm{net \thinspace flow}}$, is the change in gas density due to gas flows (independently of star formation), which can be negative, although in that case, the star formation is quenched  \citep{2019MNRAS.487L..61Z}. This term, $\dot{\overline{{\Sigma}}}_{\rm{net \thinspace flow}}$, also includes the possibility of gas return from different regions and the same region at a later epoque, an effect known as galactic fountains \citep{2017ASSL..430..323F}.
The observables are $\overline{\Sigma}_{\rm{SFR\thinspace recent}}$ and $\overline{\Sigma}_{\rm{SFR\thinspace past}}$. Let us assume that we can estimate 
$\dot{\overline{{\Sigma}}}_{\rm{gas}}$ from the star formation change considering the KS law, $\overline{\Sigma}_{\rm{SFR}}=A{\overline{{\Sigma}}}_{\rm{gas}}^N$, and rewrite Eq. \ref{eq_bath_res}:

{\small
\begin{equation} 
\begin{split}
& \overline{\Sigma}_{\rm{SFR\thinspace recent}}= \\
&  A\left\{ \left[ \dot{\overline{\Sigma}}_{\rm{net \thinspace flow}}-\overline{\Sigma}_{\rm{SFR\thinspace past}}\left( 1-R+\eta \right)   \right]\Delta t +  \left[\frac{\overline{\Sigma}_{\rm{SFR\thinspace past}}}{A} \right]^{\frac{1}{N}} \right\}^N.
\end{split}
\label{eq_etafit}
\end{equation}
}

\noindent where there is a relation between our two observables ($\overline{\Sigma}_{\rm{SFR\thinspace recent}}$ and $\overline{\Sigma}_{\rm{SFR\thinspace past}}$), $\dot{\overline{{\Sigma}}}_{\rm{net \thinspace flow}}$, and $\eta$.  We will use a simplistic approximation to estimate $\dot{\overline{{\Sigma}}}_{\rm{net \thinspace flow}}$, since we do not observe it. As explained in \cite{2019MNRAS.487L..61Z}, we assume that several regions have an approximate value close to the maximum value of $\dot{\overline{{\Sigma}}}_{\rm{net \thinspace flow}}$, for a given galaxy. 
 In the case of the estimation of $\eta$, although $\eta$ could vary between regions, we will find that the variation is smooth enough 
 (Eq. \ref{eq_eta_den}) to 
 consider the existence of a representative value for specific regions. 
  In the following, for simplicity since we are only dealing with one type of regions, the 500 pc wide ones, we will be using the analysed terms (e.g. $\overline{\Sigma}_{\rm{SFR\thinspace recent}}$, $\overline{\Sigma}_{\rm{SFR\thinspace past}}$, $\dot{\overline{{\Sigma}}}_{\rm{net \thinspace flow}}$) without the need of using the average symbols ($\Sigma_{\rm{SFR\thinspace recent}}$, $\Sigma_{\rm{SFR\thinspace past}}$, $\dot{{\Sigma}}_{\rm{net \thinspace flow}}$). Therefore, when we present an average, the average will be for several 500pc wide regions.

Assuming the instantaneous recycling approximation \citep{2014ARA&A..52..415M} 
for stars more massive than $3\rm{M_{\odot}}$ ($\tau_{\rm{MS}}\sim 0.6\rm{Gyr}$, where $\tau_{\rm{MS}}$ is the main sequence lifetime), and a Chabrier IMF 
\citep{2003PASP..115..763C}, 
we obtain a value of $R=0.27$. We use the values $A=10^{-4.32}\rm{M_{\odot}/kpc^2/yr}$, and $N=1.56$ for the KS law\cite{2007ApJ...671..333K}.

\section{Results}

\begin{figure}
%  \begin{center}
% \begin{center}
\includegraphics[width=0.5\textwidth]{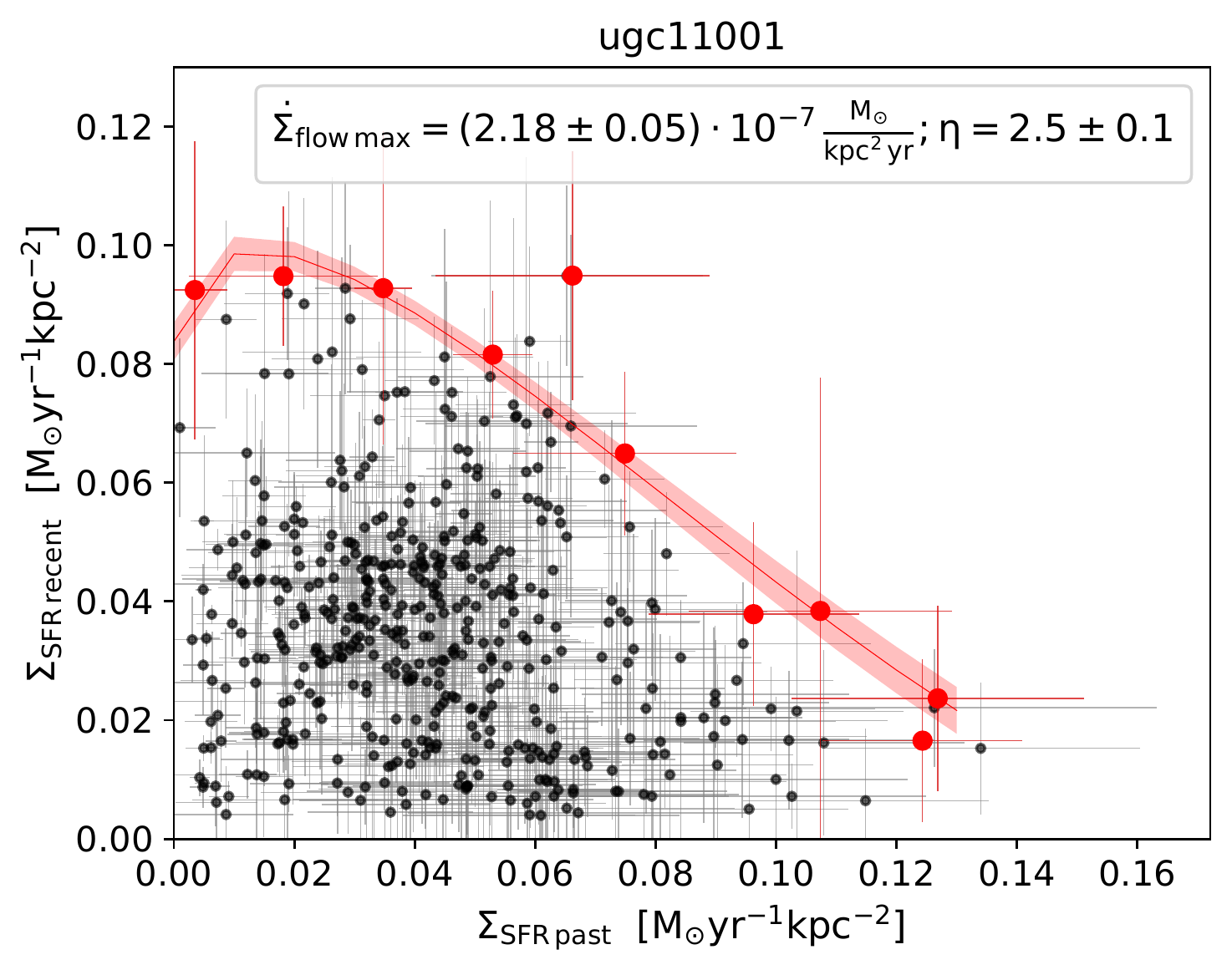} 
% \end{center}
\caption{Recent star formation rate surface density, $\Sigma_{\rm{SFR\thinspace recent}}$, versus the past 
star formation rate surface density, $\Sigma_{\rm{SFR\thinspace past}}$, for the UGC 11001 galaxy. The red dots are the regions identified as those on the envelope. We plot the fit of Eq. \ref{eq_etafit} to the regions on the envelope as well as the result of the fit and the 1-$\sigma$ uncertainty range of the fit as shaded region.
}
\label{fig_diagram_ex}
\end{figure}

As an example, we plot the $\Sigma_{\rm{SFR\thinspace recent}}$ versus $\Sigma_{\rm{SFR\thinspace past}}$ diagram for one of the galaxies, UGC 11001, in Fig. \ref{fig_diagram_ex}.  We plot the $\Sigma_{\rm{SFR\thinspace recent}}$ versus $\Sigma_{\rm{SFR\thinspace past}}$ diagrams for all of the galaxies in Fig. \ref{sfr_diagrams}. 
Each of the points in these plots can be seen as the relation between the $\Sigma_{\rm{SFR\thinspace recent}}$ and the $\Sigma_{\rm{SFR\thinspace past}}$ 
which depends on the value of $\dot{\Sigma}_{\rm{net\thinspace flow}}$, and $\eta$ (Eq. \ref{eq_etafit}). 

 We identify those regions having the maximum $\Sigma_{\rm{SFR\thinspace recent}}$, per bin of $\Sigma_{\rm{SFR\thinspace past}}$, as the regions on the envelope. 
  We see the regions on the envelope as red dots 
in Fig. \ref{fig_diagram_ex}, and as red squares in the MUSE recovered false color image of UGC 11001 in  Fig. \ref{fig_colormuse}.

\begin{table} 
% \begin{sidewaystable*}[h!]

\centering 
\caption{Estimated mass-loading factors, $\eta$, maximum flow gas surface density term, $\dot{\Sigma}_{\rm{flow\thinspace max}}$, and the associated average stellar mass surface density for the regions on the envelope, $\Sigma_{*}$.} 
\label{tab_results} 
\begin{tabular}{rrrr}  
\hline 
Galaxy identifier &  $\eta$ $^a$ & $\dot{\Sigma}_{\rm{flow\thinspace max}}$ $^b$ & $\Sigma_{*}$ $^c$ \\  &  & $10^{-8}\rm{M_{\odot}\thinspace yr^{-1}\thinspace kpc^{-2}}$ & $10^{6}\rm{M_{\odot}\thinspace kpc^{-2}}$ \\  
\hline 
pgc33816 & $4.8 \pm 0.9$ & $4.9\pm 0.9$ & $27\pm 15 $ \\ 
eso184-g082 & $5.0 \pm 1.0$ & $8.7\pm 0.6$ & $49\pm 20 $ \\ 
eso467-062 & $8.0 \pm 2.0$ & $14.0\pm 1.0$ & $51\pm 39 $ \\ 
ugc272 & $3.4 \pm 0.2$ & $6.9\pm 0.4$ & $57\pm 44 $ \\ 
ngc5584 & $2.2 \pm 0.4$ & $4.0\pm 1.0$ & $60\pm 31 $ \\ 
eso319-g015 & $5.0 \pm 2.0$ & $11.0\pm 3.0$ & $66\pm 65 $ \\ 
ugc11214 & $2.6 \pm 0.6$ & $9.0\pm 2.0$ & $84\pm 33 $ \\ 
ngc6118 & $2.19 \pm 0.09$ & $2.8\pm 0.3$ & $90\pm 51 $ \\ 
ic1158 & $6.8 \pm 0.4$ & $16.3\pm 0.3$ & $109\pm 28 $ \\ 
ngc5468 & $2.2 \pm 0.8$ & $23.0\pm 2.0$ & $113\pm 66 $ \\ 
eso325-g045 & $1.7 \pm 0.2$ & $12.0\pm 1.0$ & $121\pm 57 $ \\ 
ngc1954 & $3.3 \pm 0.5$ & $23.8\pm 0.8$ & $121\pm 31 $ \\ 
ic5332 & $3.0 \pm 1.0$ & $12.0\pm 2.0$ & $120\pm 100 $ \\ 
ugc04729 & $2.8 \pm 0.6$ & $7.0\pm 2.0$ & $126\pm 57 $ \\ 
ngc2104 & $1.7 \pm 0.6$ & $4.0\pm 2.0$ & $132\pm 62 $ \\ 
eso316-g7 & $2.0 \pm 1.0$ & $12.0\pm 6.0$ & $136\pm 39 $ \\ 
eso298-g28 & $6.0 \pm 0.7$ & $47.0\pm 2.0$ & $136\pm 46 $ \\ 
mcg-01-57-021 & $7.0 \pm 1.0$ & $17.0\pm 2.0$ & $137\pm 24 $ \\ 
pgc128348 & $2.9 \pm 0.1$ & $11.6\pm 0.7$ & $140\pm 92 $ \\ 
pgc1167400 & $2.3 \pm 0.3$ & $4.1\pm 0.8$ & $141\pm 78 $ \\ 
ngc2835 & $2.2 \pm 0.7$ & $7.0\pm 3.0$ & $144\pm 40 $ \\ 
ic2151 & $2.0 \pm 0.5$ & $5.0\pm 3.0$ & $146\pm 61 $ \\ 
ngc988 & $1.2 \pm 0.4$ & $1.0\pm 2.0$ & $158\pm 63 $ \\ 
ngc1483 & $3.0 \pm 2.0$ & $13.0\pm 5.0$ & $158\pm 40 $ \\ 
ngc7421 & $1.1 \pm 0.3$ & $1.0\pm 0.9$ & $167\pm 78 $ \\ 
fcc290 & $2.0 \pm 0.4$ & $3.0\pm 1.0$ & $169\pm 42 $ \\ 
ic344 & $2.5 \pm 0.2$ & $11.0\pm 1.0$ & $171\pm 78 $ \\ 
ngc3389 & $4.4 \pm 0.7$ & $32.4\pm 0.7$ & $190\pm 130 $ \\ 
eso246-g21 & $2.9 \pm 0.7$ & $7.0\pm 2.0$ & $188\pm 67 $ \\ 
pgc170248 & $4.9 \pm 0.7$ & $16.0\pm 1.0$ & $192\pm 77 $ \\ 
ngc7329 & $4.1 \pm 0.2$ & $7.8\pm 0.4$ & $200\pm 120 $ \\ 
ugc12859 & $2.8 \pm 0.3$ & $5.1\pm 0.9$ & $202\pm 90 $ \\ 
ugc1395 & $2.7 \pm 0.3$ & $8.0\pm 1.0$ & $200\pm 160 $ \\ 
ngc5339 & $2.9 \pm 0.4$ & $5.0\pm 1.0$ & $210\pm 150 $ \\ 
ngc1591 & $2.3 \pm 0.7$ & $19.0\pm 4.0$ & $212\pm 93 $ \\ 
pgc98793 & $1.7 \pm 0.1$ & $6.6\pm 0.8$ & $214\pm 95 $ \\ 
ugc5378 & $2.7 \pm 0.4$ & $10.0\pm 1.0$ & $223\pm 93 $ \\ 
ngc4806 & $1.9 \pm 0.2$ & $10.3\pm 0.9$ & $230\pm 170 $ \\ 
ngc1087 & $1.3 \pm 0.3$ & $14.0\pm 2.0$ & $230\pm 170 $ \\ 
ngc4980 & $1.8 \pm 0.3$ & $6.6\pm 0.9$ & $240\pm 170 $ \\ 
ngc6902 & $2.4 \pm 0.3$ & $8.0\pm 1.0$ & $240\pm 13 $ \\ 
ugc11001 & $2.5 \pm 0.1$ & $21.8\pm 0.5$ & $250\pm 140 $ \\ 
ic217 & $1.2 \pm 0.4$ & $3.0\pm 2.0$ & $266\pm 88 $ \\ 
eso506-g004 & $3.4 \pm 0.2$ & $13.9\pm 0.4$ & $270\pm 170 $ \\ 
ic2160 & $2.4 \pm 0.4$ & $12.0\pm 2.0$ & $270\pm 260 $ \\ 
ngc1385 & $0.9 \pm 0.1$ & $9.7\pm 0.9$ & $272\pm 44 $ \\ 
mcg-01-33-034 & $1.0 \pm 0.09$ & $7.9\pm 0.5$ & $270\pm 150 $ \\ 
ngc4603 & $0.9 \pm 0.1$ & $3.0\pm 1.0$ & $276\pm 85 $ \\ 
ngc4535 & $4.1 \pm 0.7$ & $15.0\pm 2.0$ & $280\pm 200 $ \\ 
ngc1762 & $2.3 \pm 0.2$ & $8.2\pm 0.7$ & $290\pm 140 $ \\ 
ngc3451 & $3.7 \pm 0.4$ & $11.0\pm 1.0$ & $300\pm 180 $ \\ 
ngc4790 & $1.2 \pm 0.3$ & $9.0\pm 2.0$ & $334\pm 63 $ \\ 
ngc3244 & $1.8 \pm 0.2$ & $8.0\pm 1.0$ & $340\pm 270 $ \\ 
ngc628 & $1.8 \pm 0.1$ & $19.0\pm 1.0$ & $360\pm 170 $ \\ 
pgc30591 & $0.7 \pm 0.4$ & $0.0\pm 2.0$ & $360\pm 160 $ \\ 
ngc5643 & $1.1 \pm 0.3$ & $3.0\pm 2.0$ & $410\pm 160 $ \\ 
ngc1309 & $1.3 \pm 0.2$ & $17.0\pm 2.0$ & $420\pm 220 $ \\ 

\end{tabular} 
\\
\hrulefill
\\
{\raggedright 
$^a$ Mass-loading factor derived in this work.\  
$^b$ Maximum flow gas surface density term derived in this work.\  
$^c$ Stellar mass surface density for the regions on the envelope obtained in this work. }  
\\
\hrulefill
\end{table} 
\begin{table} 
% \begin{sidewaystable*}[h!]

\centering 
\contcaption{}
\begin{tabular}{rrrr}  
\hline 
Galaxy identifier &  $\eta$ $^a$ & $\dot{\Sigma}_{\rm{flow\thinspace max}}$ $^b$ & $\Sigma_{*}$ $^c$ \\  &  & $10^{-8}\rm{M_{\odot}\thinspace yr^{-1}\thinspace kpc^{-2}}$ & $10^{6}\rm{M_{\odot}\thinspace kpc^{-2}}$ \\  
\hline 
ngc1084 & $0.6 \pm 0.2$ & $13.0\pm 6.0$ & $420\pm 170 $ \\ 
ngc7580 & $1.6 \pm 0.2$ & $18.0\pm 1.0$ & $420\pm 110 $ \\ 
ngc692 & $2.7 \pm 0.2$ & $14.0\pm 1.0$ & $420\pm 220 $ \\ 
eso462-g009 & $4.0 \pm 1.0$ & $8.0\pm 2.0$ & $440\pm 310 $ \\ 
ic5273 & $1.5 \pm 0.7$ & $9.0\pm 4.0$ & $450\pm 220 $ \\ 
pgc3140 & $1.0 \pm 0.1$ & $4.0\pm 1.0$ & $460\pm 210 $ \\ 
ic1553 & $0.6 \pm 0.3$ & $5.0\pm 2.0$ & $460\pm 290 $ \\ 
ugc11289 & $1.8 \pm 0.3$ & $8.0\pm 3.0$ & $472\pm 46 $ \\ 
ic4582 & $2.0 \pm 0.1$ & $9.1\pm 0.5$ & $480\pm 280 $ \\ 
ngc2466 & $1.5 \pm 0.2$ & $18.0\pm 2.0$ & $480\pm 470 $ \\ 
eso443-21 & $1.9 \pm 0.3$ & $94.0\pm 4.0$ & $490\pm 180 $ \\ 
ic4452 & $0.37 \pm 0.05$ & $5.2\pm 0.6$ & $500\pm 320 $ \\ 
eso498-g5 & $1.5 \pm 0.3$ & $20.0\pm 2.0$ & $510\pm 320 $ \\ 
eso552-g40 & $1.7 \pm 0.4$ & $12.0\pm 3.0$ & $540\pm 280 $ \\ 
eso163-g11 & $0.48 \pm 0.06$ & $1.0\pm 2.0$ & $570\pm 300 $ \\ 
ngc7582 & $0.6 \pm 0.1$ & $0.0\pm 2.0$ & $570\pm 280 $ \\ 
ngc1620 & $1.4 \pm 0.3$ & $8.0\pm 2.0$ & $580\pm 400 $ \\ 
ic1320 & $1.4 \pm 0.1$ & $4.6\pm 0.9$ & $590\pm 270 $ \\ 
ngc3393 & $1.9 \pm 0.3$ & $11.0\pm 1.0$ & $590\pm 270 $ \\ 
ngc2370 & $0.87 \pm 0.08$ & $4.0\pm 2.0$ & $600\pm 310 $ \\ 
ngc4981 & $0.85 \pm 0.08$ & $7.0\pm 1.0$ & $630\pm 580 $ \\ 
ngc3783 & $1.3 \pm 0.3$ & $12.0\pm 3.0$ & $680\pm 170 $ \\ 
ngc1285 & $0.8 \pm 0.1$ & $7.0\pm 2.0$ & $700\pm 460 $ \\ 
ngc5806 & $1.7 \pm 0.3$ & $17.0\pm 3.0$ & $720\pm 530 $ \\ 
eso018-g018 & $1.0 \pm 0.3$ & $13.0\pm 5.0$ & $720\pm 420 $ \\ 
ngc6754 & $0.44 \pm 0.06$ & $1.0\pm 1.0$ & $750\pm 480 $ \\ 
ic2560 & $1.2 \pm 0.2$ & $11.0\pm 1.0$ & $760\pm 370 $ \\ 
ngc7140 & $2.4 \pm 0.7$ & $13.0\pm 6.0$ & $770\pm 210 $ \\ 
ngc3464 & $1.9 \pm 0.3$ & $9.0\pm 3.0$ & $780\pm 380 $ \\ 
mcg-02-13-38 & $1.3 \pm 0.1$ & $20.0\pm 1.0$ & $790\pm 450 $ \\ 
ngc1590 & $0.83 \pm 0.08$ & $25.0\pm 3.0$ & $810\pm 460 $ \\ 
pgc8822 & $1.15 \pm 0.09$ & $4.0\pm 1.0$ & $970\pm 430 $ \\ 
ngc7721 & $1.3 \pm 0.5$ & $10\pm 10$ & $990\pm 540 $ \\ 
pgc28308 & $0.2 \pm 0.1$ & $0.0\pm 8.0$ & $1030\pm 450 $ \\ 
ngc1137 & $0.8 \pm 0.2$ & $1.0\pm 4.0$ & $1030\pm 830 $ \\ 
eso478-g006 & $1.1 \pm 0.2$ & $10.0\pm 4.0$ & $1050\pm 650 $ \\ 
ngc1448 & $1.05 \pm 0.09$ & $15.0\pm 2.0$ & $1130\pm 480 $ \\ 
ngc3278 & $1.3 \pm 0.2$ & $31.0\pm 3.0$ & $1170\pm 710 $ \\ 
ngc4030 & $0.6 \pm 0.1$ & $14.0\pm 3.0$ & $1310\pm 550 $ \\ 
ngc3363 & $0.41 \pm 0.08$ & $0.0\pm 0.9$ & $1330\pm 590 $ \\ 
ngc7780 & $1.2 \pm 0.2$ & $8.0\pm 3.0$ & $1450\pm 590 $ \\ 
ic1438 & $1.5 \pm 0.3$ & $11.0\pm 4.0$ & $1520\pm 250 $ \\ 
ngc4666 & $0.3 \pm 0.1$ & $10.0\pm 3.0$ & $2900\pm 1300 $ \\ 
ngc7396 & $0.5 \pm 0.2$ & $0.0\pm 4.0$ & $4100\pm 1900 $ \\ 
ngc716 & $0.5 \pm 0.2$ & $70\pm 10$ & $5200\pm 1900 $ \\ 
\hline 
\end{tabular} 

\end{table} 

We can see in Eq. \ref{eq_etafit} that we need the value of $\dot{\Sigma}_{\rm{net \thinspace flow}}$ in combination with the $\Sigma_{\rm{SFR\thinspace recent}}$ and $\Sigma_{\rm{SFR\thinspace past}}$ values, in order to quantify $\eta$. 
For a given galaxy, there should be a maximum value for the net flow gas surface density term, $\dot{\Sigma}_{\rm{flow\thinspace max}}$.  Although in principle $\dot{\Sigma}_{\rm{flow\thinspace max}}$ is unknown for us, if we assume that there are several segments where $\dot{\Sigma}_{\rm{net\thinspace flow}}\sim\dot{\Sigma}_{\rm{flow\thinspace max}}$, then these regions are those on the $\Sigma_{\rm{SFR\thinspace recent}}$ versus $\Sigma_{\rm{SFR\thinspace past}}$ diagram envelope.

 Assuming $\dot{\Sigma}_{\rm{net\thinspace flow}}$ constant, we fit Eq. \ref{eq_etafit} to the regions on the envelope and estimate the mass-loading factor, representative of those specific regions. 
 We have selected a galaxy sample mainly composed by galaxies on the main sequence of star formation, and remove segments having very high $\Sigma_{\rm{SFR\thinspace recent}}$ compared to the rest of the segments of a galaxy.
 Therefore, the galaxy sample, as well as the segments, have been chosen to assure the $\dot{\Sigma}_{\rm{net\thinspace flow}}\sim\dot{\Sigma}_{\rm{flow\thinspace max}}$ hypothesis for several regions. 
 The regions below the envelope are due, to a greater extent,  to regions having a smaller value 
of $\dot{\Sigma}_{\rm{net\thinspace flow}}<\dot{\Sigma}_{\rm{flow\thinspace max}}$, and to a much lesser
extent, to regions having different $\eta$ values.
 
We have fitted Eq. \ref{eq_etafit} to the envelopes of the 102 galaxy listed in Table \ref{tab_sample} and show the results in Fig. \ref{fig_diagram_ex} for UGC 11001, and in Fig. \ref{sfr_diagrams} for the rest of the galaxies, as a red line.

\subsection{ Variations of $\eta$ }

\begin{figure}
\begin{center}
\includegraphics[width=0.49\textwidth]{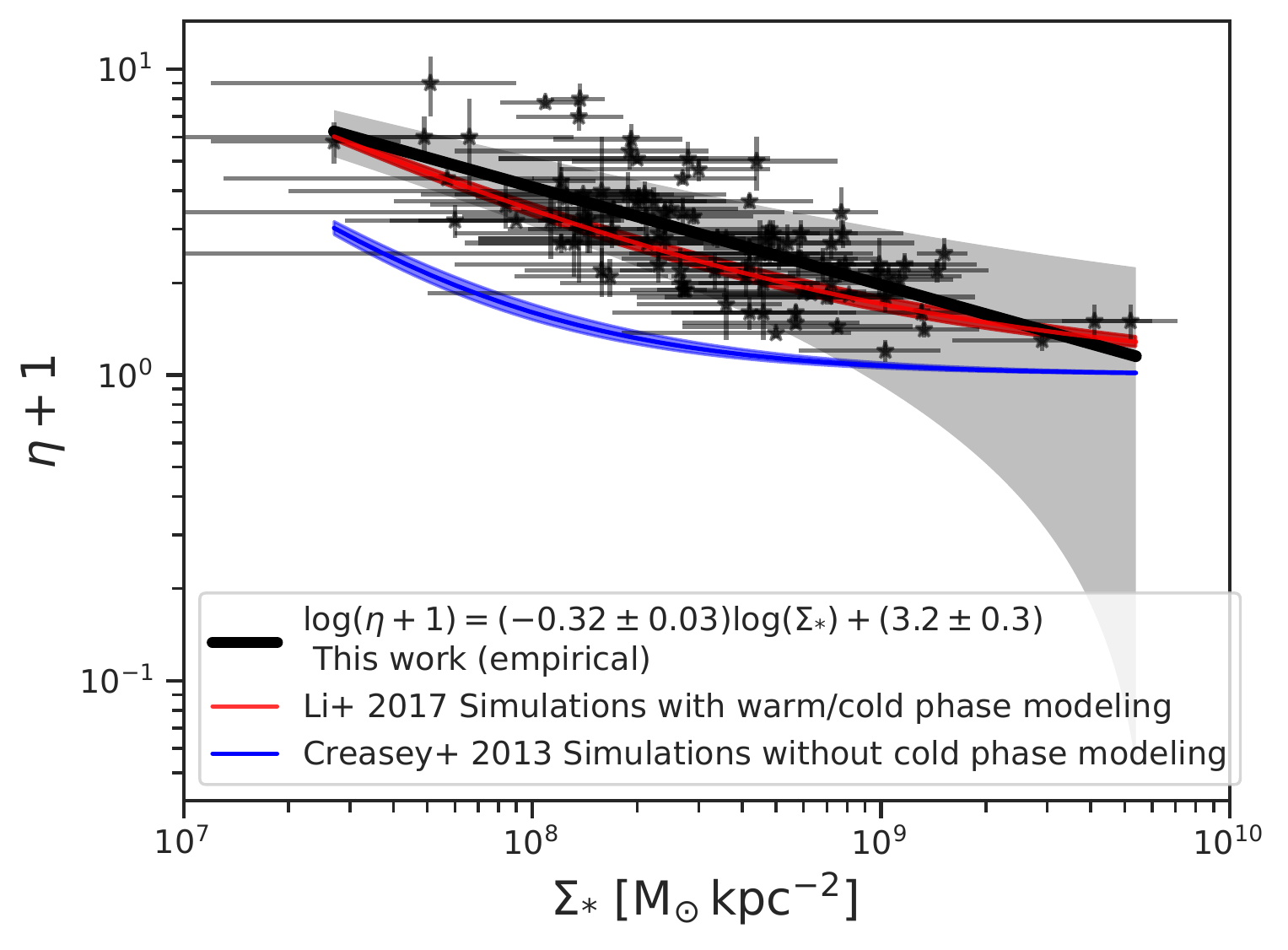} 
\end{center}
\caption{Local mass-loading factor, $\eta$, versus the average stellar mass surface density 
of the regions on the envelope for each galaxy, $\Sigma_{*}$, as black stars. We plot the linear fit to the empirically derived quantities and the 1-$\sigma$ uncertainty range of the fit as shaded regions.
The blue line shows the $\eta$ reported for supernova explosions hydrodynamical simulations \citep{2013MNRAS.429.1922C} with cut-off cooling modeling, and as blue line the $\eta$ reported for supernova explosions hydrodynamical 
simulations  \citep{2017ApJ...841..101L} with gas cold-phase modeling included.
}
\label{fig_eta_den}
\end{figure}

Although we have an $\eta$ value for each galaxy, $\eta$ is an average value representative only of the regions on the envelope, instead of the whole galaxy. We associate the estimated $\eta$ for a given envelope with the average surface stellar mass density, $\Sigma_{*}$, of those regions on the envelope where we estimate $\eta$. Therefore, $\eta$ is a local average value, representative only of the regions on the envelope, and their mean value of $\Sigma_{*}$.
Although $\eta$ value might vary through the regions on the envelope, we assume that the variation is smooth enough and associate the average stellar mass 
density, $\Sigma_{*}$ and the standard deviation, to each envelope.  The correlation found between $\eta$ and $\Sigma_{*}$ (Eq. \ref{eq_eta_den} and Fig. \ref{fig_eta_den}) is in fact smooth enough to make the association between $\eta$ and $\Sigma_{*}$ for the regions on the envelope.  
We report $\eta$, $\dot{\Sigma}_{\rm{flow\thinspace max}}$, and $\Sigma_{*}$ values in Table \ref{tab_results}.

We plot in Fig. \ref{fig_eta_den}, $\eta$ versus $\Sigma_{*}$, and find that the mass-loading factors strongly correlate with the local $\Sigma_{*}$ measured on the envelope regions: 

\begin{equation}
 \log(\eta+1)=(-0.32\pm0.03)\log(\Sigma_{*})+(3.2\pm0.3)
 \label{eq_eta_den}
\end{equation}

This correlation means that the denser the region, the lower is the mass-loading factor, which means that  the amount of outflowing gas mass per unit star formation rate depends inversely on the stellar mass surface density. This is so because the denser the region, the larger is the local gravitational pull, making it harder for the gas to be expelled. The local chemical enrichment of galaxies \citep{2018ApJ...852...74B} also favours the gas regulator model and finds that the mass-loading factor depends on the local escape velocity. 
This new empirical $\eta$-$\Sigma_{*}$ relation, can be used to check if stellar feedback implementations in numerical simulations  \citep{2014MNRAS.445..581H,2017ApJ...841..101L,2013MNRAS.429.1922C} are realistic. In particular, we can 
compare our $\eta$-$\Sigma_{*}$ relation with the $\eta$-$\Sigma_{\rm{gas}}$ relation from supernova explosion feedback simulations \citep{2017ApJ...841..101L,2013MNRAS.429.1922C},
using the observed $\Sigma_{*}$-$\Sigma_{\rm{gas}}$ relation \citep{2020MNRAS.492.2651B}. The inverse correlation between $\eta$ and $\Sigma_{*}$ is clear from observations and simulations. However, our empirical result supports that the radiative cooling below $10^{4}\rm{K}$ is important. This is because when ignoring gas cooling below $10^4\rm{K}$, there is an excess of warm gas compared with the models including a multiphase cold/warm gas  \citep{2017ApJ...841..101L}.
Thus, in the warm gas excess scenario,  there is a layer of gas with higher ISM pressure which prevents the gas to be expelled from the supernova explosions.

\section{Local to global mass-loading factors}

The mass-loading factor derived here 
is representative of local scales. However, other observational and theoretical studies report global 
mass-loading factors  \citep{2015MNRAS.454.2691M,2016MNRAS.455.2592R,2017MNRAS.465.1682H,2019MNRAS.490.4368S,2019ApJ...886...74M}. We estimate global mass-loading factors, $\eta_{\rm{G}}$, from the empirical 
$\eta$-$\Sigma_{*}$ relation reported here (Eq. \ref{eq_eta_den}), integrating over observed stellar mass density profiles. 

To convert $\eta$ to $\eta_{\rm{G}}$, we assume that we can estimate the total outflow due to stellar feedback, $\dot{M}_{\rm{out}}$, by adding up $\dot{\Sigma}_{\rm{out}}$ over each individual segment where stellar feedback acts. Since we are interested in galaxy discs, we assume a radial characterization for the properties of interest, i.e., $\eta$,  
$\dot{\Sigma}_{\rm{out}}$, $\Sigma_{\rm{SFR}}$, and $\Sigma_{\rm{SFR}}$ depend on $R$, the radial distance to the centre of the disc: 
\begin{equation}
 \eta_{\rm{G}}=\frac{\dot{M}_{\rm{out}}}{\rm{SFR}}=\frac{\int_{0}^{\infty}\dot{\Sigma}_{\rm{out}}(R)R\thinspace{\rm{d}}R}{\int_{0}^{\infty} \Sigma_{\rm{SFR}}(R)R\thinspace{\rm{d}}R}=\frac{\int_{0}^{\infty}\eta(R) \Sigma_{\rm{SFR}}(R)R\thinspace{\rm{d}}R}{\int_{0}^{\infty} \Sigma_{\rm{SFR}}(R)R\thinspace{\rm{d}}R}.
 \label{eta_global}
\end{equation}

The conversion between $\eta$ to $\eta_{\rm{G}}$ is a $\Sigma_{\rm{SFR}}$ weighted average of $\eta$. Stellar mass density profiles, $\Sigma_{*}(R)$, are better constrained than $\Sigma_{\rm{SFR}}(R)$ profiles, so we decide to use the empirical relation between 
$\Sigma_{\rm{SFR}}\propto \Sigma_{*}(R)^n$. For simplicity, we assume $n=1$, which is consistent with the latest results of this relation for galaxy discs \citep{2019MNRAS.488.3929C}, although we found that different $n$ values close to 1 do not change the results significantly. Using our empirical relation (Eq. \ref{eq_eta_den}) we rewrite Eq. \ref{eta_global}: 

\begin{equation}
  \eta_{\rm{G}}=\frac{\int_{0}^{\infty}\eta(R) \Sigma_{*}(R)R\thinspace{\rm{d}}R}{\int_{0}^{\infty} \Sigma_{*}(R)R\thinspace{\rm{d}}R}=\frac{10^{3.2}\int_{0}^{\infty}{\Sigma_{*}(R)}^{0.68} R\thinspace{\rm{d}}R}{\int_{0}^{\infty} \Sigma_{*}(R)R\thinspace{\rm{d}}R}.
 \label{eta_global2}
\end{equation}

The stellar mass surface density profile, $\Sigma_{*}(R)$, is therefore all we need to compute the global mass-loading factor. We use the deepest stellar mass surface density profiles from the Spitzer Survey of Stellar Structure in Galaxies (S4G)   \citep{2016A&A...596A..84D} to compute the global mass-loading factor as a function of stellar mass. S4G stellar mass surface density profiles are divided into 5 mass bins: $[10^{8.5}$-$10^{9}]$, $[10^{9}$-$10^{9.5}]$, $[10^{9.5}$-$10^{10}]$, $[10^{10}$-$10^{10.5}]$, and $[10^{10.5}$-$10^{11}]\thinspace\rm{M_{\odot}}$. We used these 5 mass bins to compute the $\eta_{\rm{G}}$ shown in Fig. \ref{eta_mstar}.

\subsection{Variations of $\eta_{\rm{G}}$}
\begin{figure}
\begin{center}
\includegraphics[width=0.49\textwidth]{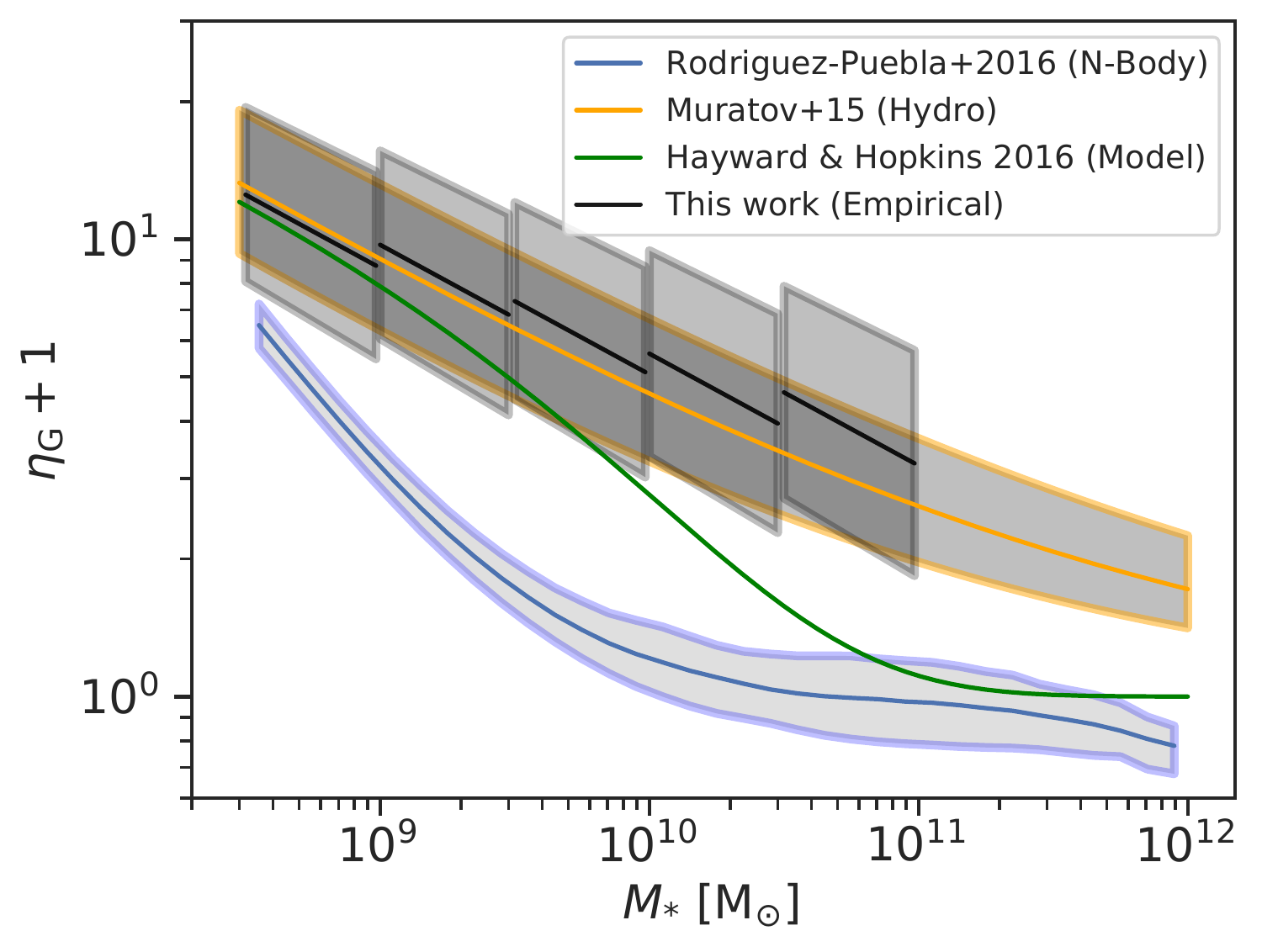} 
\end{center}
\caption{Global mass-loading factor, $\eta_{\rm{G}}$, versus the stellar mass of the galaxy, $M_*$. 
The black lines show the empirical quantities which have been derived combining the distribution of $\eta$ as a function of stellar mass surface density (Fig. \ref{fig_eta_den}) with the stellar mass surface density profiles from S4G results  \citep{2016A&A...596A..84D}, where the shaded region is the corresponding uncertainty propagated from the 1-$\sigma$ uncertainty found in the $\eta$-$\Sigma_{*}$ relation (Eq. \ref{eq_eta_den}). The discontinuity is due to the division in different parametrizations of the stellar mass surface density profiles  in 5  mass bins reported by \citet{2016A&A...596A..84D}. The orange line shows the $\eta_{\rm{G}}$ reported for cosmological zoom-in galaxy hydrodynamical simulations \citep{2015MNRAS.454.2691M}, the green line the shows an analytical feedback model \citep{2017MNRAS.465.1682H}, and the blue line shows the $\eta_{\rm{G}}$ reported using the N-body Bolshoi-Planck simulation \citep{2016MNRAS.455.2592R}.
}
\label{eta_mstar}
\end{figure}

We present our 
empirically derived global mass-loading factors as a function of stellar mass in Fig. \ref{eta_mstar}  as black lines. The discontinuity is due to the division in different parametrizations of the stellar mass surface density profiles  in 5  mass bins reported by \citet{2016A&A...596A..84D} . 
The most important feature we find in Fig. \ref{eta_mstar} is that the smaller the stellar mass of the galaxy, the larger the mass-loading factor, as required to reconcile the ratio of halo to stellar mass in low-mass galaxies \citep{2010ApJ...717..379B,2016MNRAS.455.2592R}. We compare our empirical estimates with predictions based on N-body \citep{2016MNRAS.455.2592R} and hydrodynamical simulations \citep{2015MNRAS.454.2691M}, as well as with an analytical model \citep{2017MNRAS.465.1682H}.

Some simulations, as well as the analytic model we used to compare with, define the mass-loading factor using the outflowing mass that escapes the galaxy forever \citep{2016MNRAS.455.2592R,2017MNRAS.465.1682H}. Therefore, they  
do not consider the outflowing gas which returns to the galaxy at a later time which,  depending on the outflow velocity, will result in smaller mass-loading factors. We define the mass-loading factor as the total mass ejected 
independently of velocity, and the mass can come back at a later time (outside the 550Myr time range). There is a  remarkable agreement between our 
empirical result and that from hydrodynamical simulations that define the mass-loading factor as independent of velocity  \citep{2015MNRAS.454.2691M}. In fact, for lower masses ($M_* < 10^{9.5}\rm{M_{\odot}}$), where the 
stellar feedback is thought to be more important regulating galaxy stellar mass growth, and the escape velocity lower, the agreement with the analytic model is also remarkably good. 

\section{Discussion}

\subsection{Comparison with other studies}
The results reported here are slightly different when we compare them with 
recent observational studies reporting local \citep{2019Natur.569..519K,2020MNRAS.493.3081R} and global \citep{2019MNRAS.490.4368S,2019ApJ...886...74M} mass-loading factors. 

The reported local mass-loading factors using the spatial de-correlation between star formation and molecular gas \citep{2019Natur.569..519K} differ by less than $3\sigma$ from our reported values, and it might be  due to the smaller time scale of $\sim 1.5\rm{Myr}$ for which they report efficient gas dispersal, while our reported time scale is $\sim550\rm{Myr}$. 
The previously reported local mass-loading factors using the Na {\sc{D}} absorption \citep{2020MNRAS.493.3081R} are consistent with ours within $1\sigma$.

The use of Mg {\sc{II}} absorption of the circum-galactic medium to derive global mass-loading factors gives no
clear dependence on the total mass of the galaxy \citep{2019MNRAS.490.4368S}. However, the Mg {\sc{II}} absorption method gives $\eta_{\rm{G}}$ with very
high uncertainties, mainly due to the uncertainty when deriving the HI
column density from the  Mg {\sc{II}} equivalent width  \citep{2015ApJ...804...83S}. Due to these large uncertainties, their results are apparently consistent
with our results within 1$\sigma$ for most of their reported 
$\eta_{\rm{G}}$'s.

The mass-loading factor estimates using deep H$\alpha$ imaging give smaller mass-loading factors compared to those reported here and give no correlation with the stellar mass of the galaxy \citep{2019ApJ...886...74M}. Nevertheless, the method using H$\alpha$ imaging derives the amount of outflowing gas from the H$\alpha$ surface brightness background, so it neglects H$\alpha$ emission stronger than this background emission. 
The estimated outflowing mass could be inferior to the real one since we already know that H$\alpha$ emission has a component due to expansive bubbles  \citep{2005A&A...430..911R,2015MNRAS.447.3840C}. 

\subsection{Discussion on envelope's shapes}

Eq. \ref{eq_etafit} depends on $\dot{\Sigma}_{\rm{flow \thinspace max}}$, $\eta$, $\Sigma_{\rm{SFR\thinspace recent}}$, and $\Sigma_{\rm{SFR\thinspace past}}$ values. The case shown in Fig. \ref{fig_diagram_ex}, where there is no high recent star formation rate for those regions where the past star formation rate was the highest, is a common case, but not the only one. For instance, there are cases where $\Sigma_{\rm{SFR\thinspace recent}}$ values are approximately constant through the regions on the envelope or even increase as   
$\Sigma_{\rm{SFR\thinspace past}}$ increase (e.g. NGC 988, NGC 7421, IC 217, IC 4452, PGC 28308 in Fig. \ref{sfr_diagrams}). 

Essentially, there are two terms depending on $\Sigma_{\rm{SFR\thinspace past}}$ in Eq. \ref{eq_etafit}, 
one with a direct proportionality and the other with an inverse one. 
The latter dominates for larger $\eta$ values meaning that the larger the mass-loading factor, the larger is the effect in reducing the amount of gas to form stars, as expected. However, for small enough $\eta$ and $\Sigma_{\rm{SFR\thinspace past}}$ values, the directly dependent term can dominate producing a direct relation between $\Sigma_{\rm{SFR\thinspace recent}}$ and $\Sigma_{\rm{SFR\thinspace past}}$, as we see in some galaxies in Fig. \ref{sfr_diagrams} (e.g. IC 4452).
The extrapolation of Eq. \ref{eq_etafit} to large enough values of $\Sigma_{\rm{SFR\thinspace past}}$ would give always a decrease in $\Sigma_{\rm{SFR\thinspace recent}}$, as long as $\eta>0$, and that is the reason to observe some inverted U-shape envelopes (e.g. NGC 1084, PGC 3140). 

Finally, for large enough $\dot{\Sigma}_{\rm{flow \thinspace max}}$ values, the directly $\Sigma_{\rm{SFR\thinspace past}}$ dependent term is almost negligible for small values of $\Sigma_{\rm{SFR\thinspace past}}$, and then the recent star formation depends on $\dot{\Sigma}_{\rm{flow \thinspace max}}$ for the low $\Sigma_{\rm{SFR\thinspace past}}$, while decreases proportionally with $\Sigma_{\rm{SFR\thinspace past}}$ depending on the mass-loading factor (e.g  UGC 5378, UGC 11001).

Therefore, the combination of $\dot{\Sigma}_{\rm{flow \thinspace max}}$ and $\eta$ variations, as well as the range covered by the past and recent star formation rate surface densities, is what gives the different envelope's shapes.  
The uncertainties obtained when fitting Eq. \ref{eq_etafit} show reliable estimates, except for one case, the PGC 30591 galaxy, which is the one having the highest  inclination of our sample.

\subsection{High-inclination galaxies}

Although we have removed edge-on ($i=90^{\circ}$) galaxies from our sample, high-inclination ($i>70^{\circ}$) galaxies might not be good candidates to apply the method used in this study, as in the case of the PGC 30591 galaxy. In the case of edge-on and very high-inclination galaxies, two main effects can affect the pertinence of applying the method. Firstly, one can have in the line of sight a large combination of different regions of the galaxy. Secondly, the higher the inclination, the smaller is the number of resolved regions, while the method relies on a large enough number of resolved regions. 

\begin{figure}
\begin{center}
\includegraphics[width=0.49\textwidth]{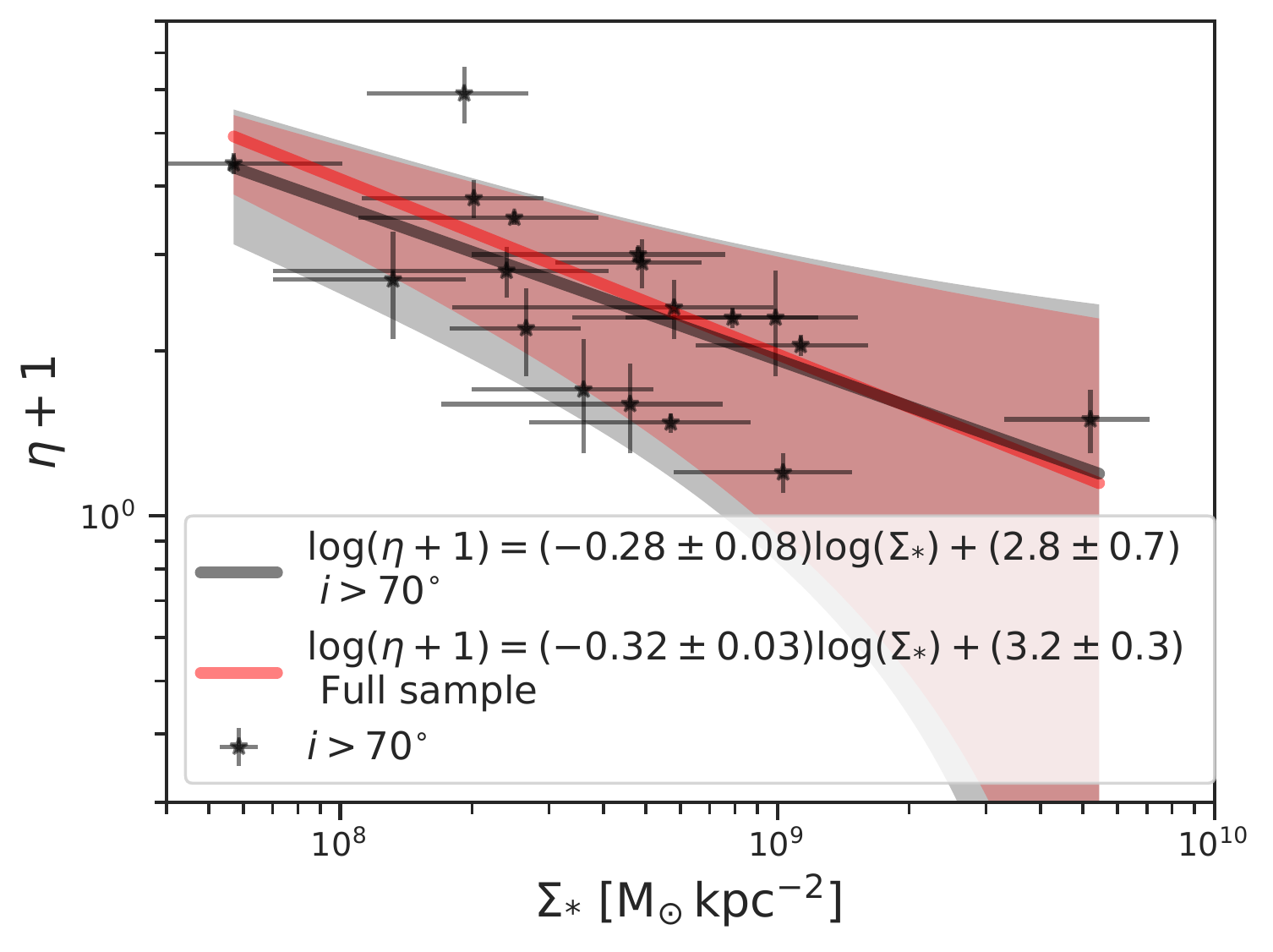} 
\end{center}
\caption{Local mass-loading factor, $\eta$, versus the average stellar mass surface density 
of the regions on the envelope for each galaxy, $\Sigma_{*}$, as black stars, for high-inclination galaxies. We plot, as a black line, the linear fit to the empirically derived quantities.
We plot as a red line, the fit for the full sample (Eq. \ref{eq_eta_den}), and the 1-$\sigma$ uncertainty range of the fits as shaded regions.
}
\label{eta_highincl}
\end{figure}

Nevertheless, except for PGC 30591, we found reliable $\dot{\Sigma}_{\rm{flow \thinspace max}}$ and $\eta$ estimates, even for 
high-inclination galaxies. Including or excluding high-inclination galaxies does not make any changes to the reported $\eta$-$\Sigma_{*}$ relation presented here. We show in Fig. \ref{eta_highincl} the $\eta$-$\Sigma_{*}$ observed and fitted relations for high-inclination galaxies, as star symbols and black line, respectively. When we compare the high-inclination galaxies results with the fit obtained using the full sample (Eq. \ref{eq_eta_den}), shown as a red line, both samples are compatible within $1\sigma$.

\subsection{Effects of very high recent SFR regions removal}

\begin{figure}
\begin{center}
\includegraphics[width=0.49\textwidth]{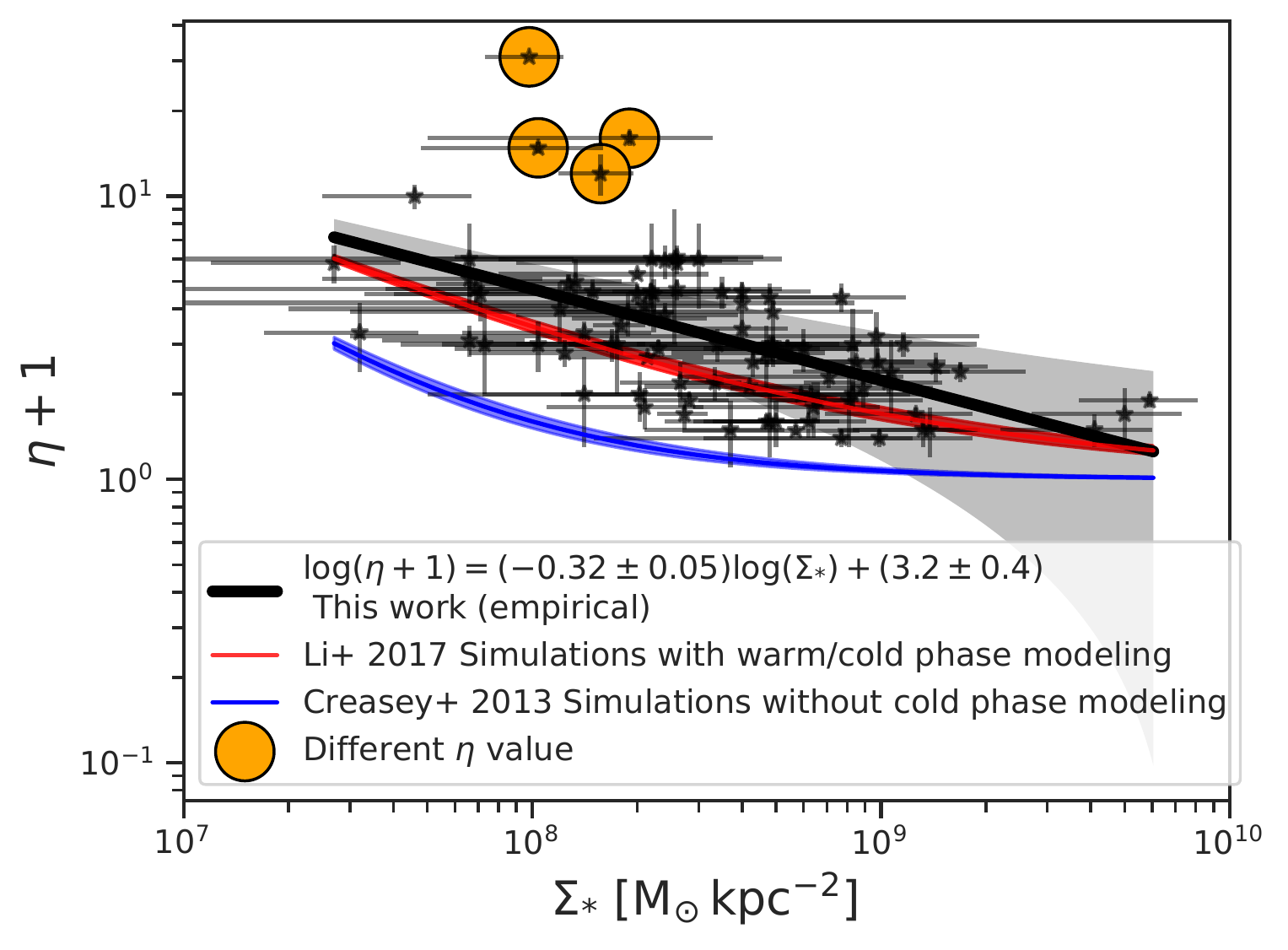} 
\end{center}
\caption{ Local mass-loading factor, $\eta$, versus the average stellar mass surface density 
of the regions on the envelope for each galaxy, $\Sigma_{*}$, as black stars. Same plot as in Fig. \ref{fig_eta_den}, but this version of the plot has been done without removing any region based on its $\Sigma_{\rm{SFR\thinspace recent}}$ value. The 4 galaxies where we find a significant difference in the derived $\eta$ value are marked as orange circles.
We plot the linear fit to the empirically derived quantities and the 1-$\sigma$ uncertainty range of the fit as shaded regions.
The blue line shows the $\eta$ reported for supernova explosions hydrodynamical simulations \citep{2013MNRAS.429.1922C} with cut-off cooling modeling, and as blue line the $\eta$ reported for supernova explosions hydrodynamical 
simulations  \citep{2017ApJ...841..101L} with gas cold-phase modeling included.
}
\label{fig_eta_den_rem}
\end{figure}

We think is important to remove regions having  very high values of $\Sigma_{\rm{SFR\thinspace recent}}$ compared to the rest of the galaxy, since these regions probably have a very high $\dot{\Sigma}_{\rm{net\thinspace flow}}$ value compared to the rest of the regions identified to be on the envelope. The effect of considering these high $\Sigma_{\rm{SFR\thinspace recent}}$ regions affects our assumption about the approximate equal value of $\dot{\Sigma}_{\rm{net\thinspace flow}}$ for the regions on the envelope.  

However, in order to explore the effect of this removal in our results, we have performed the same analysis without removing any region based on its $\Sigma_{\rm{SFR\thinspace recent}}$ value. We plot the derived $\eta$ versus the local $\Sigma_{*}$ of the regions on the envelope for the whole sample of galaxies in Fig. \ref{fig_eta_den_rem}. We also plot the resulted $\eta$-$\Sigma_{*}$ fit to the observed data. We find a very similar correlation between $\eta$ and the local $\Sigma_{*}$:

\begin{equation}
 \log(\eta+1)=(-0.32\pm0.05)\log(\Sigma_{*})+(3.2\pm0.4)
 \label{eq_eta_den_rem}.
\end{equation}

There are only 4 galaxies where the $\eta$ value significantly changes: IC 1158, NGC 3389, ESO 298-G28, and MCG 01-57-021. We marked these galaxies as orange circles in Fig. \ref{fig_eta_den_rem}. However, the changes are not significant enough to change the resulted $\eta$-$\Sigma_{*}$ relation, but just a slightly increase in the obtained uncertainties. Therefore, although the removal could be important for some specific individual galaxies, when compared with the full sample of galaxies we still find a consistent $\eta$-$\Sigma_{*}$ relation.

\section{Conclusions}

We have used MUSE observations of a sample of 102 galaxy discs. We extracted 
the spectra of 500 pc wide regions and apply them stellar population synthesis 
using SINOPSIS code. We obtained the star formation histories of those regions 
and we analysed the recent and past star formation rate densities. We compared   
the $\Sigma_{\rm{SFR\thinspace recent}}$ with the $\Sigma_{\rm{SFR\thinspace past}}$ and 
found that, for each galaxy, there is an envelope of regions formed by those regions having the maximum $\Sigma_{\rm{SFR\thinspace recent}}$, per bin of $\Sigma_{\rm{SFR\thinspace past}}$. 
We fitted the  resolved star formation self-regulator model (Eq. \ref{eq_etafit}) to those regions on the envelope 
and quantify the  
mass-loading factor, $\eta$.

 We find correlations locally between $\eta$ and the stellar mass surface density, $\Sigma_{*}$, and globally between the averaged value of $\eta$ for a galaxy, $\eta_{\rm{G}}$, and the stellar mass of the galaxy, $M_{*}$, which are strong indications of how stellar feedback {\bf locally} regulates the mass growth of galaxies, especially those of lower masses.
  The comparison between our empirical local $\eta$-$\Sigma_{*}$ relation with that from hydrodynamical simulations of supernova explosions \citep{2017ApJ...841..101L} is remarkably in agreement. In the case of our empirical global $\eta_{\rm{G}}$-$M_{*}$ relation, the 
 comparison with hydrodynamical cosmological zoom-in galaxy simulations \citep{2015MNRAS.454.2691M} is also in excellent agreement.
 
 We note that the value of $\eta$ depends on the time scale over which the feedback is analysed, and can be defined either including or excluding posterior gas return, so comparison with other observations and with theory must be done with care.
 These empirical relations offer excellent tools to confront with stellar feedback models which are crucial for understanding galaxy formation and evolution.

\section*{Acknowledgements}
The authors thank the anonymous referee whose comments have  led  to  significant  improvements  in  the  paper. The authors also thank Aldo Rodr\'iguez-Puebla for sharing their stellar halo accretion rate coevolution models from \citet{2016MNRAS.455.2592R}.
JZC and IA's work is funded by a CONACYT grant through project FDC-2018-1848. 
DRG acknowledges financial support through CONACYT project  A1-S-22784. 
  GB acknowledges financial support through PAPIIT project IG100319 from DGAPA-UNAM.
 This research has made use of the services of the ESO Science Archive Facility, 
 Astropy,\footnote{\url{http://www.astropy.org}} a 
 community-developed core Python package for Astronomy \citep{2013A&A...558A..33A,2018AJ....156..123A}, APLpy, 
 an open-source plotting package for Python  \citep{2012ascl.soft08017R}, Astroquery, a package that contains a collection of tools to access online Astronomical data \citep{2019AJ....157...98G}, and pyregion ({\url{https://github.com/astropy/pyregion}), a python module to parse ds9 region files.
 We acknowledge the usage of the HyperLeda database ({\url{http://leda.univ-lyon1.fr}}).
 Based on observations collected at the European Southern Observatory 
 under ESO programmes 095.D-0172(A), 196.B-0578(B), 096.D-0263(A), 097.D-0408(A), 095.B-0532(A), 096.D-0296(A), 60.A-9319(A), 1100.B-0651(B), 0102.B-0794(A), 098.B-0551(A), 0101.D-0748(A), 099.B-0397(A), 1100.B-0651(A), 0100.D-0341(A), 096.B-0309(A), 296.B-5054(A), 095.D-0091(B), 0101.D-0748(B), 095.D-0091(A), 097.B-0640(A), 096.B-0054(A), 1100.B-0651(C), 098.D-0115(A), 097.B-0165(A), 098.C-0484(A), 094.C-0623(A), 60.A-9339(A), 099.B-0242(A), 0102.A-0135(A), 097.B-0041(A), 095.B-0934(A), 099.D-0022(A), and 60.A-9329(A).

%%%%%%%%%%%%%%%%%%%%%%%%%%%%%%%%%%%%%%%%%%%%%%%%%%

%%%%%%%%%%%%%%%%%%%% REFERENCES %%%%%%%%%%%%%%%%%%

% The best way to enter references is to use BibTeX:

\bibliographystyle{mnras}
% \bibliography{SR_javier_v1} % if your bibtex file is called example.bib

 \section*{Data Availability}
The data underlying this article are available in the ESO 
archive, at \url{http://archive.eso.org/}. The datasets were derived for each source through the query form at \url{http://archive.eso.org/wdb/wdb/adp/phase3_spectral/form?collection_name=MUSE}.

% Alternatively you could enter them by hand, like this:
% This method is tedious and prone to error if you have lots of references
% \begin{thebibliography}{99}
% 
% % 
% \end{thebibliography}

%%%%%%%%%%%%%%%%%%%%%%%%%%%%%%%%%%%%%%%%%%%%%%%%%%

%%%%%%%%%%%%%%%%% APPENDICES %%%%%%%%%%%%%%%%%%%%%

% \clearpage
% \end{document}
\appendix
\section{SFR recent-past diagrams}

\onecolumn
\begin{figure} 
\subfloat{ \includegraphics[width=0.3\linewidth]{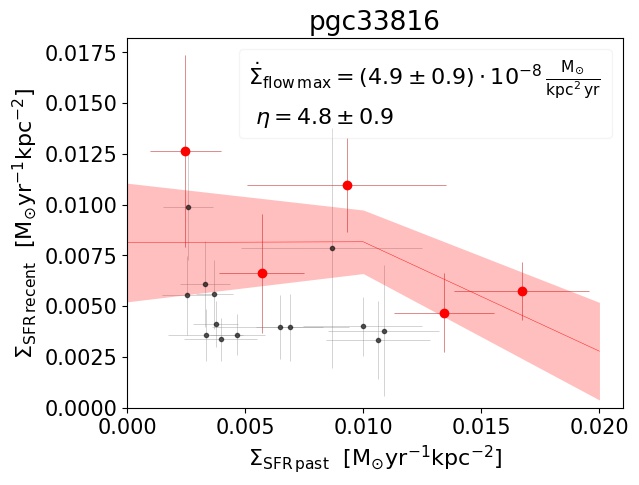}}  
 \subfloat{ \includegraphics[width=0.3\linewidth]{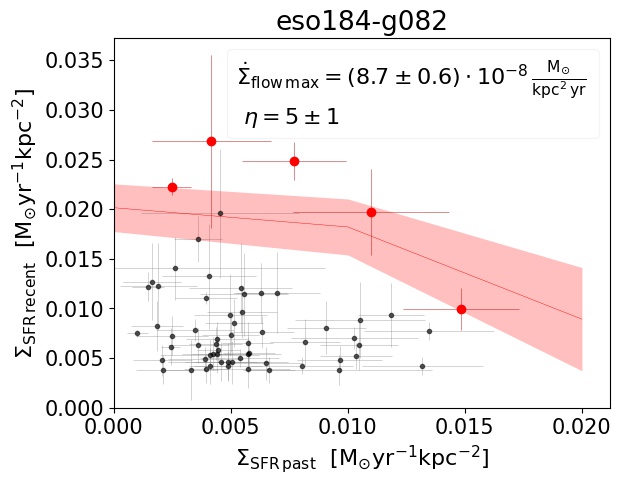}}  
 \subfloat{ \includegraphics[width=0.3\linewidth]{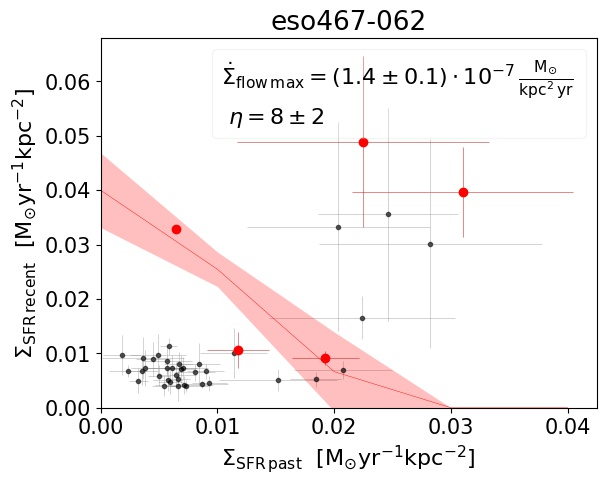}}  \\ 
\subfloat{ \includegraphics[width=0.3\linewidth]{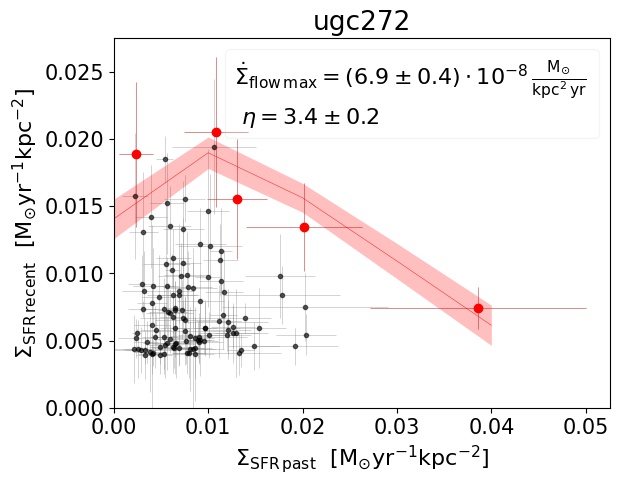}}  
 \subfloat{ \includegraphics[width=0.3\linewidth]{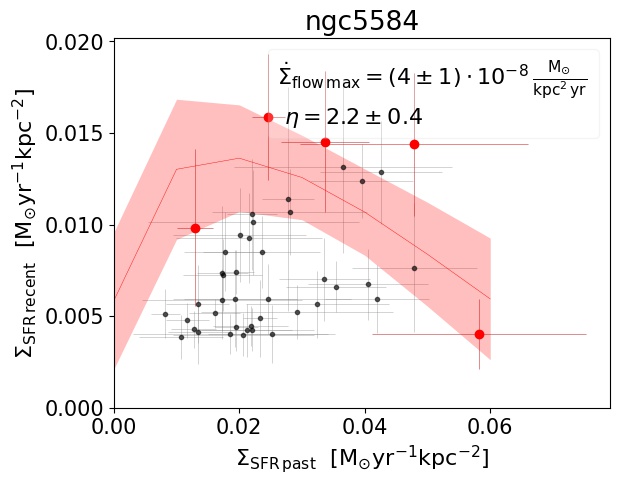}}  
 \subfloat{ \includegraphics[width=0.3\linewidth]{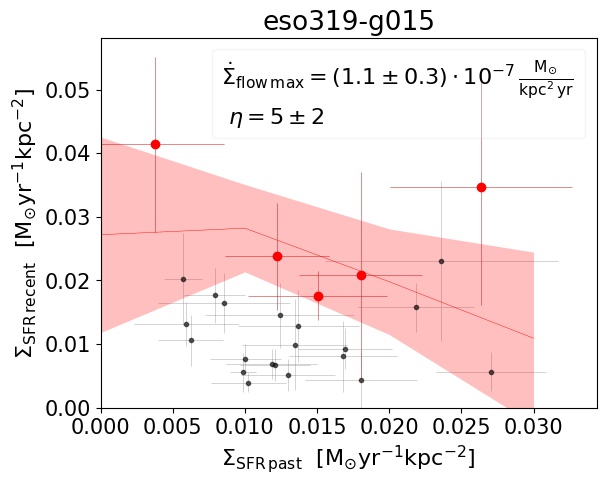}}  \\ 
\subfloat{ \includegraphics[width=0.3\linewidth]{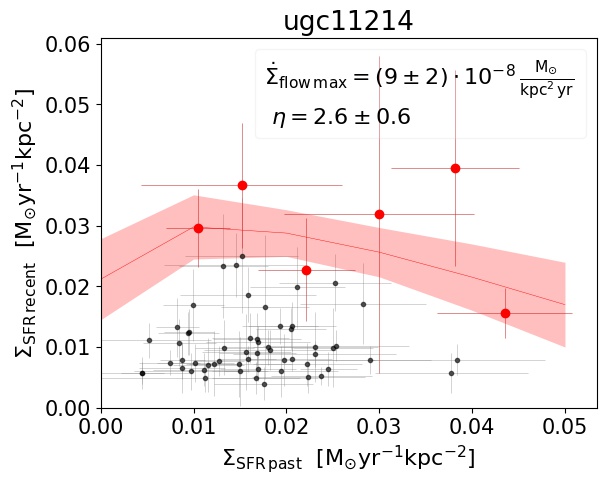}}  
 \subfloat{ \includegraphics[width=0.3\linewidth]{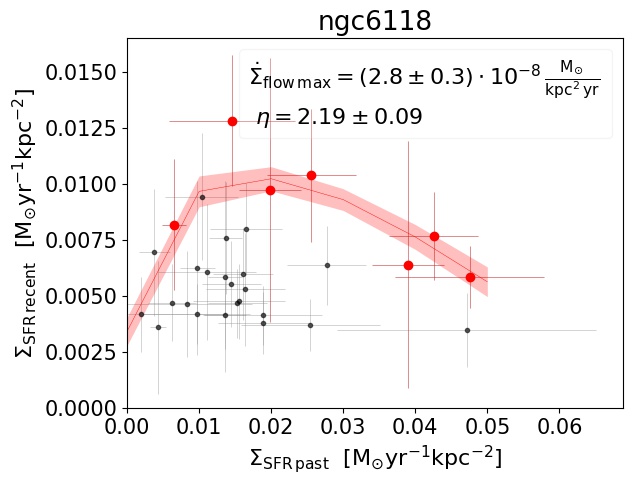}}  
 \subfloat{ \includegraphics[width=0.3\linewidth]{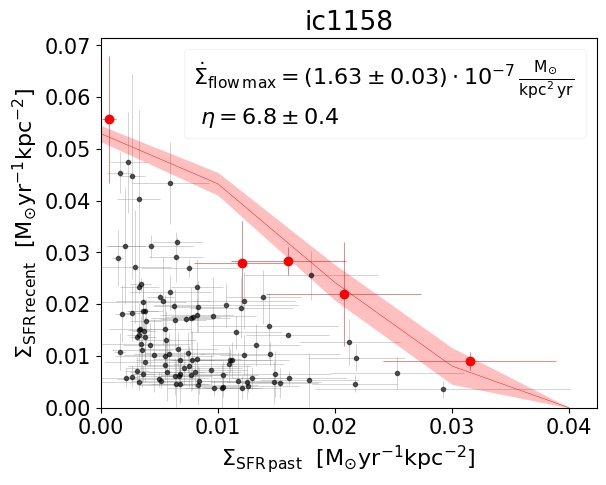}}  \\ 
\subfloat{ \includegraphics[width=0.3\linewidth]{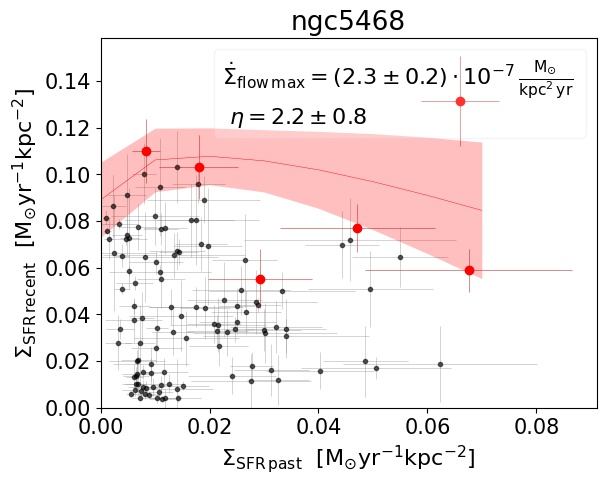}}  
 \subfloat{ \includegraphics[width=0.3\linewidth]{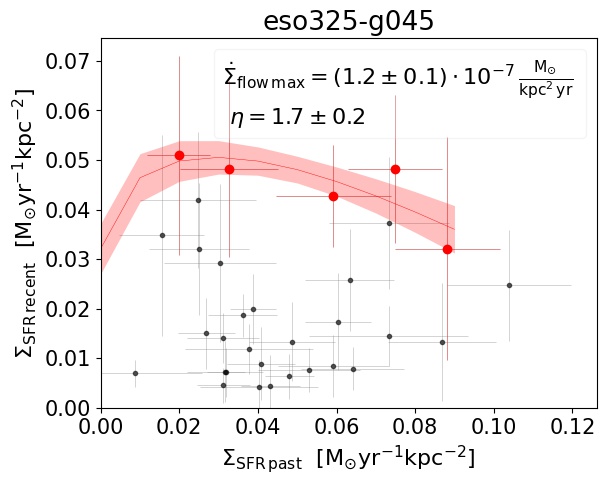}}  
 \subfloat{ \includegraphics[width=0.3\linewidth]{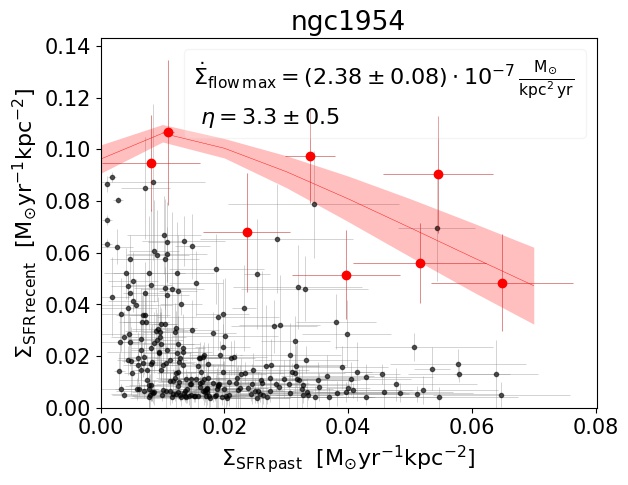}}  \\ 
\caption{\label{sfr_diagrams}  {\bf SFR recent-past diagrams}. Recent star formation surface density, $\Sigma_{\rm{SFR\thinspace recent}}$, versus the past star formation rate surface density, $\Sigma_{\rm{SFR\thinspace past}}$, for each galaxy of the sample. The red dots are the regions identified as those on the envelope. We plot the fit of Eq. \ref{eq_etafit} to the regions on the envelope and the 1-$\sigma$ uncertainty range of the fit as shadow region. We show the parameters of the fit, at the top of each panel.} 
\end{figure} 
\begin{figure} 
\subfloat{ \includegraphics[width=0.3\linewidth]{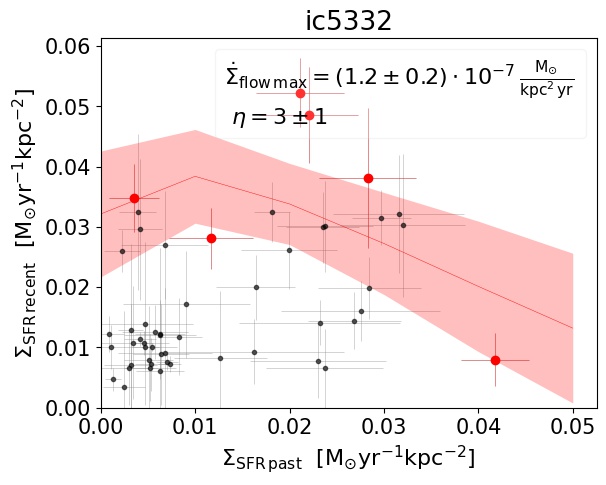}}  
 \subfloat{ \includegraphics[width=0.3\linewidth]{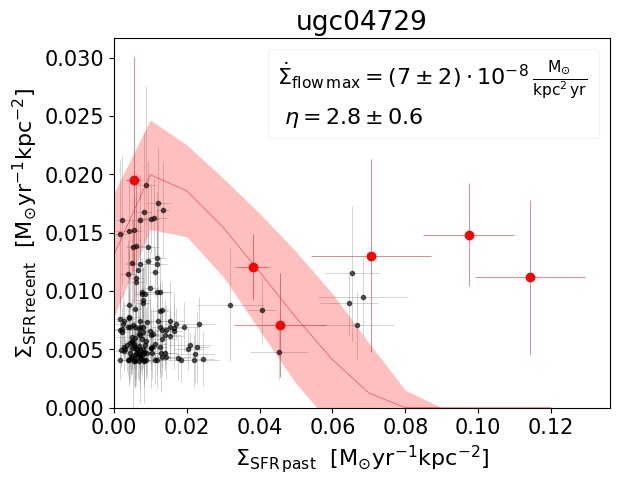}}  
 \subfloat{ \includegraphics[width=0.3\linewidth]{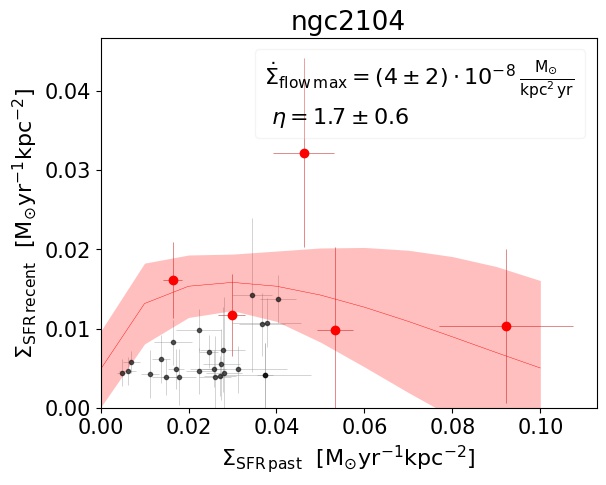}}  \\ 
\subfloat{ \includegraphics[width=0.3\linewidth]{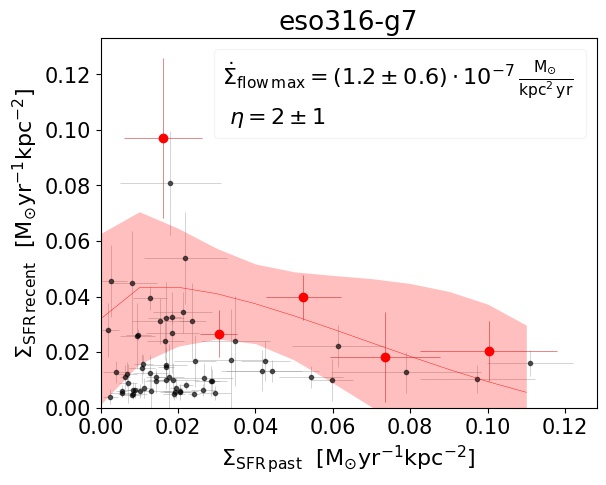}}  
 \subfloat{ \includegraphics[width=0.3\linewidth]{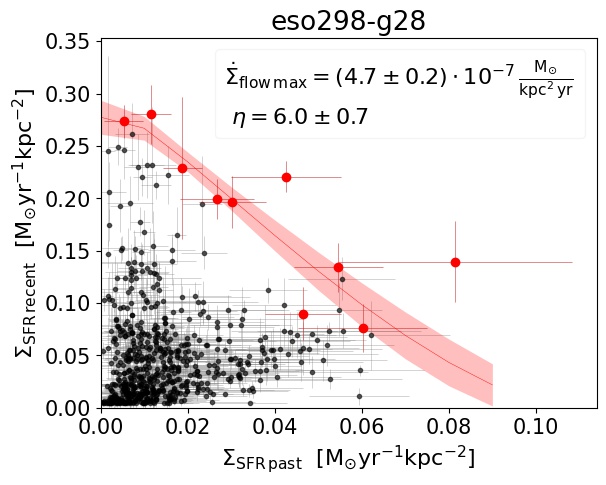}}  
 \subfloat{ \includegraphics[width=0.3\linewidth]{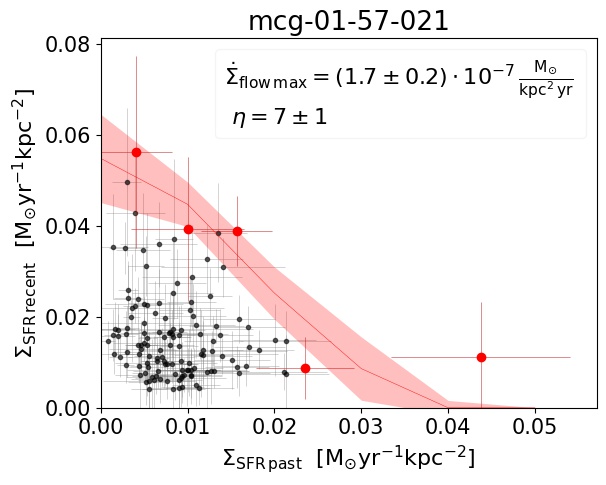}}  \\ 
\subfloat{ \includegraphics[width=0.3\linewidth]{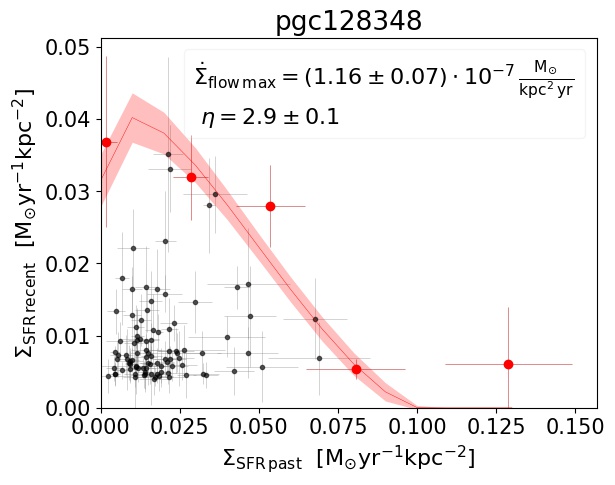}}  
 \subfloat{ \includegraphics[width=0.3\linewidth]{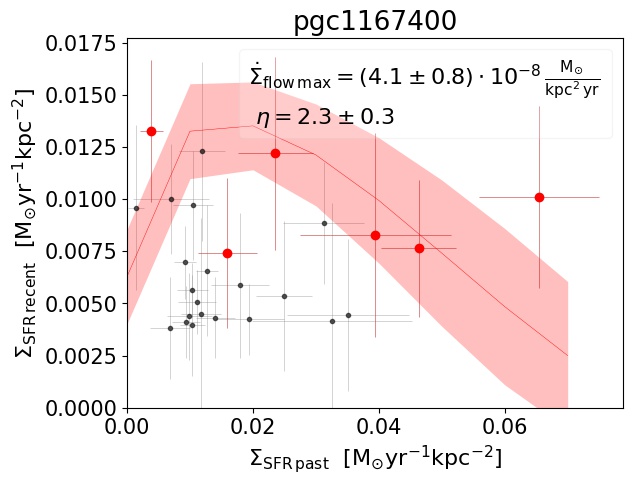}}  
 \subfloat{ \includegraphics[width=0.3\linewidth]{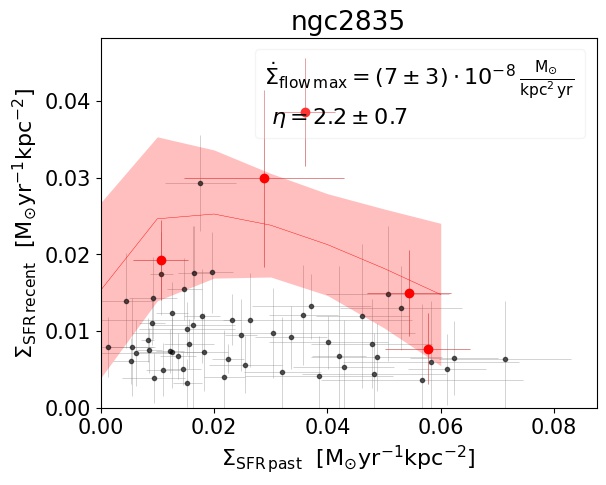}}  \\ 
\subfloat{ \includegraphics[width=0.3\linewidth]{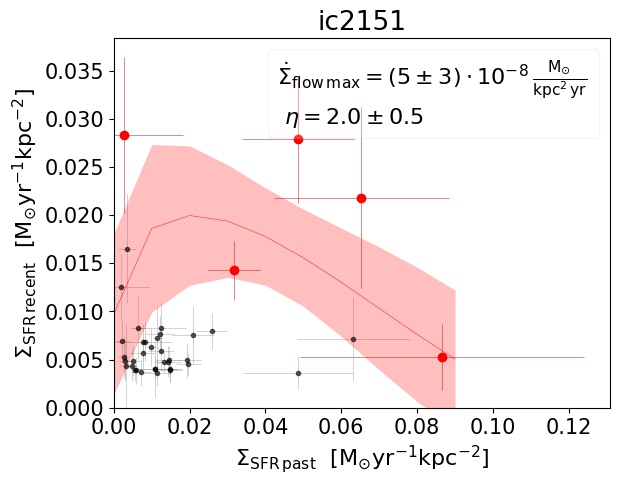}}  
 \subfloat{ \includegraphics[width=0.3\linewidth]{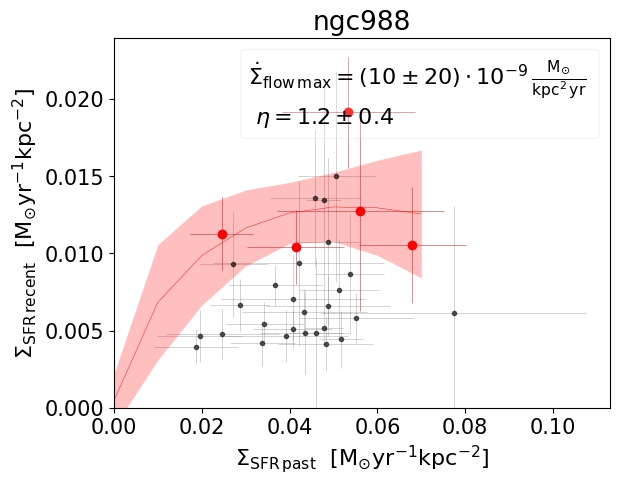}}  
 \subfloat{ \includegraphics[width=0.3\linewidth]{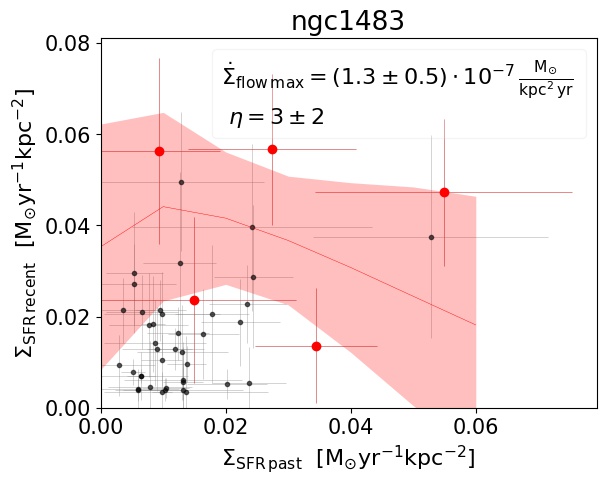}}  \\ 
\contcaption{} 
\end{figure} 
\begin{figure} 
\subfloat{ \includegraphics[width=0.3\linewidth]{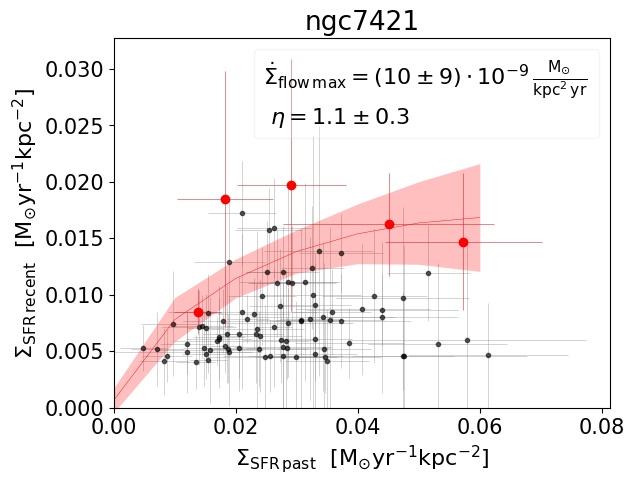}}  
 \subfloat{ \includegraphics[width=0.3\linewidth]{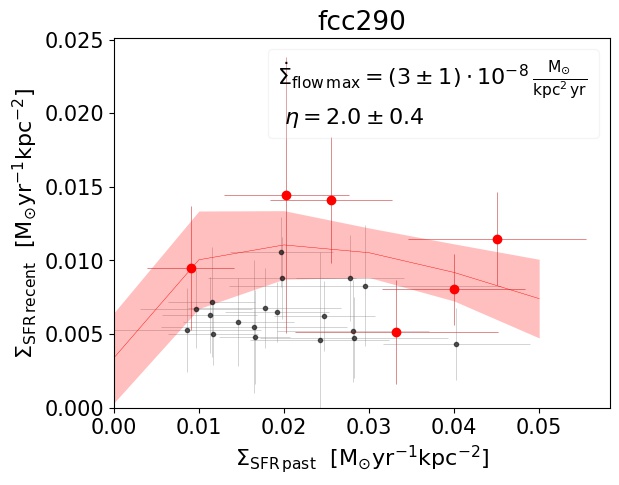}}  
 \subfloat{ \includegraphics[width=0.3\linewidth]{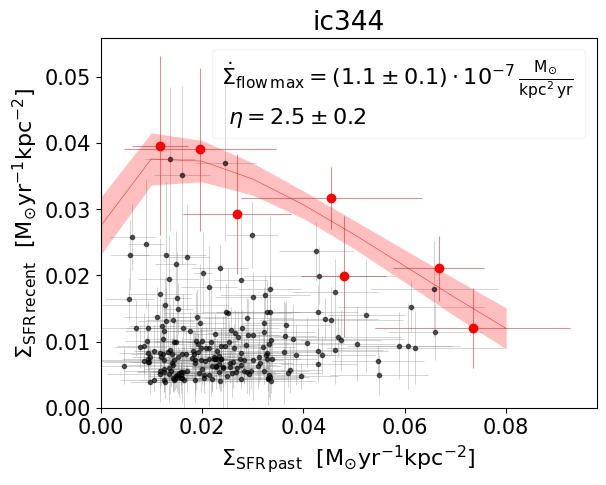}}  \\ 
\subfloat{ \includegraphics[width=0.3\linewidth]{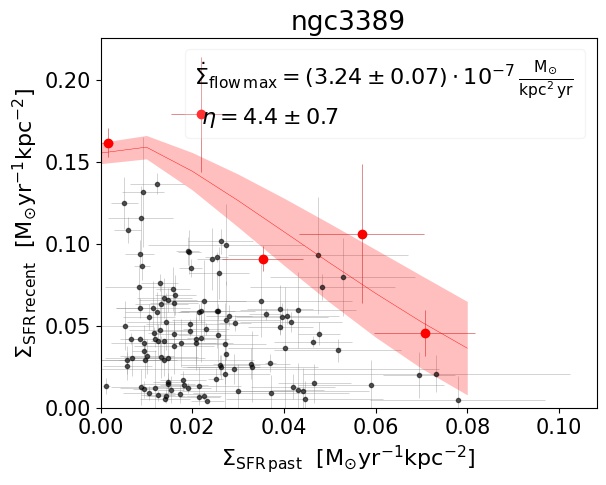}}  
 \subfloat{ \includegraphics[width=0.3\linewidth]{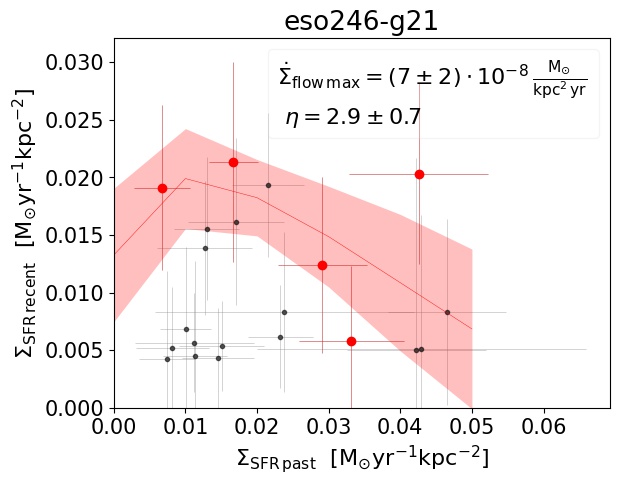}}  
 \subfloat{ \includegraphics[width=0.3\linewidth]{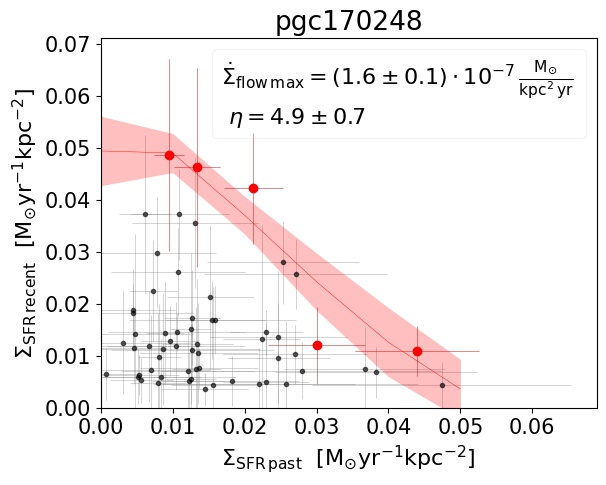}}  \\ 
\subfloat{ \includegraphics[width=0.3\linewidth]{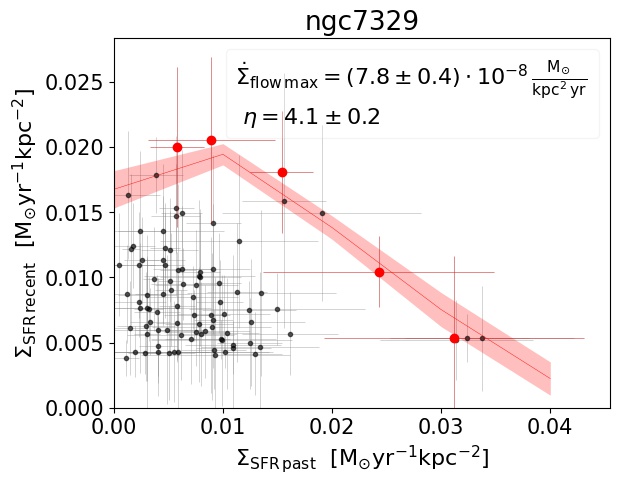}}  
 \subfloat{ \includegraphics[width=0.3\linewidth]{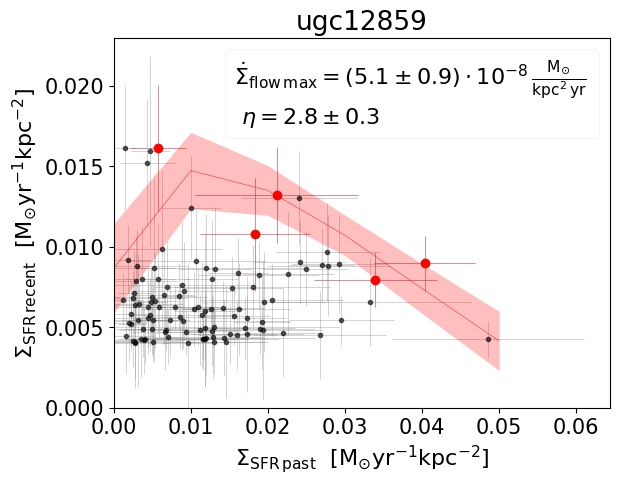}}  
 \subfloat{ \includegraphics[width=0.3\linewidth]{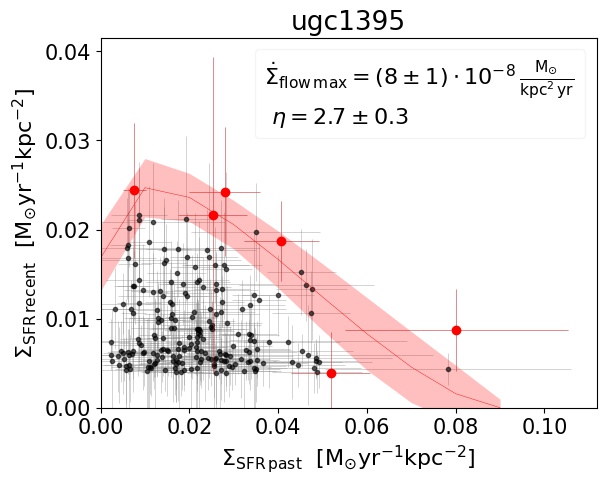}}  \\ 
\subfloat{ \includegraphics[width=0.3\linewidth]{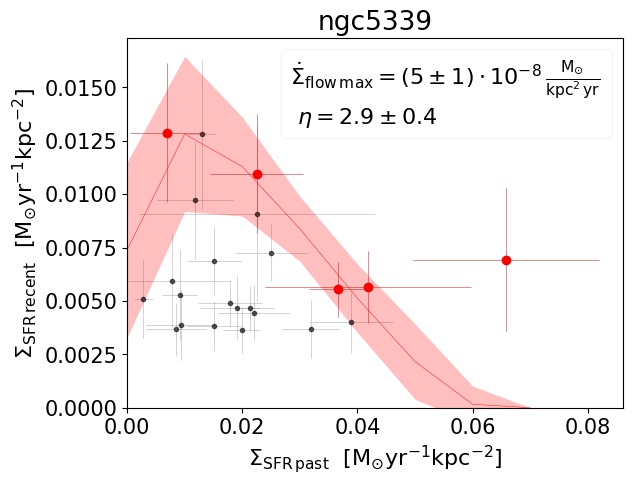}}  
 \subfloat{ \includegraphics[width=0.3\linewidth]{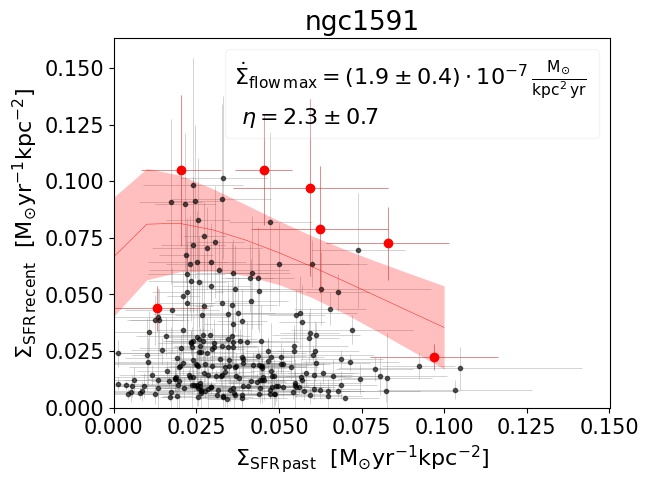}}  
 \subfloat{ \includegraphics[width=0.3\linewidth]{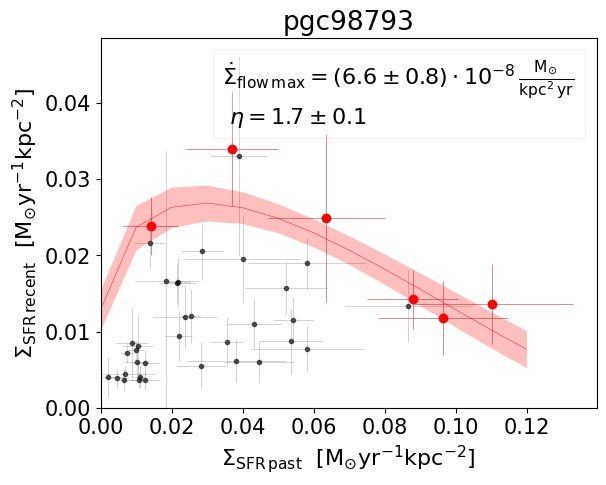}}  \\ 
\contcaption{} 
\end{figure} 
\begin{figure} 
\subfloat{ \includegraphics[width=0.3\linewidth]{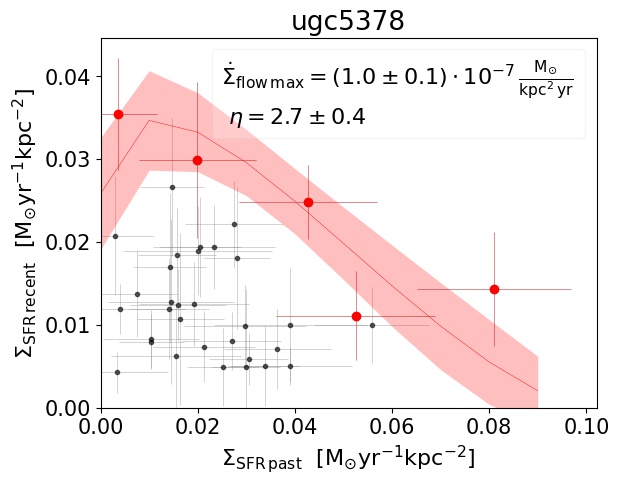}}  
 \subfloat{ \includegraphics[width=0.3\linewidth]{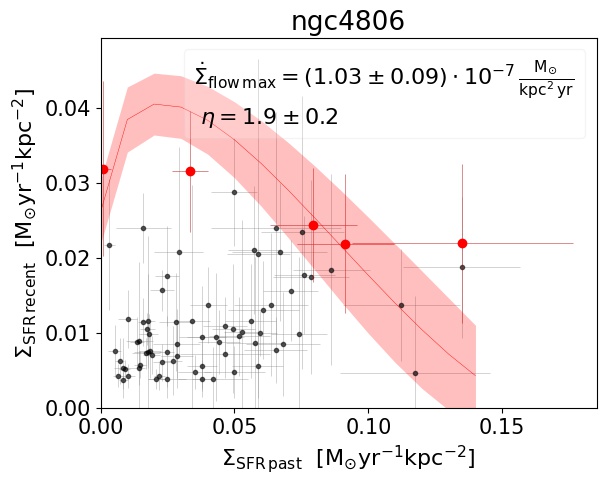}}  
 \subfloat{ \includegraphics[width=0.3\linewidth]{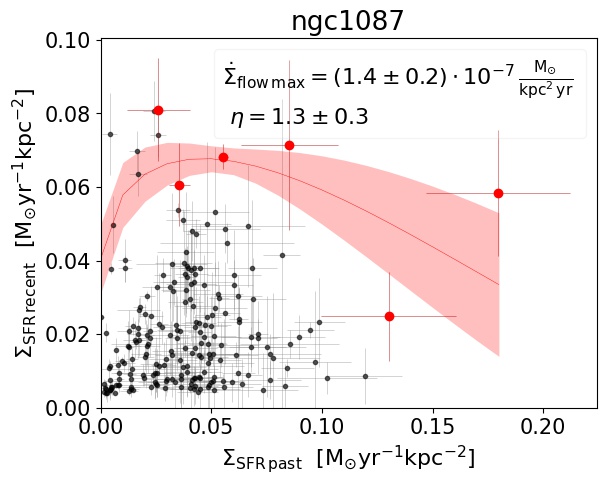}}  \\ 
\subfloat{ \includegraphics[width=0.3\linewidth]{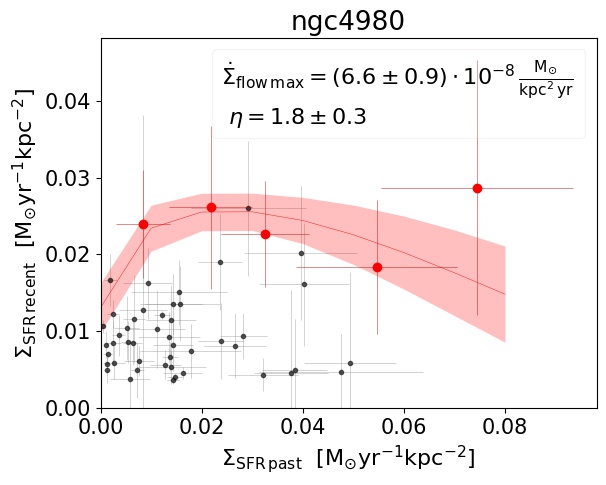}}  
 \subfloat{ \includegraphics[width=0.3\linewidth]{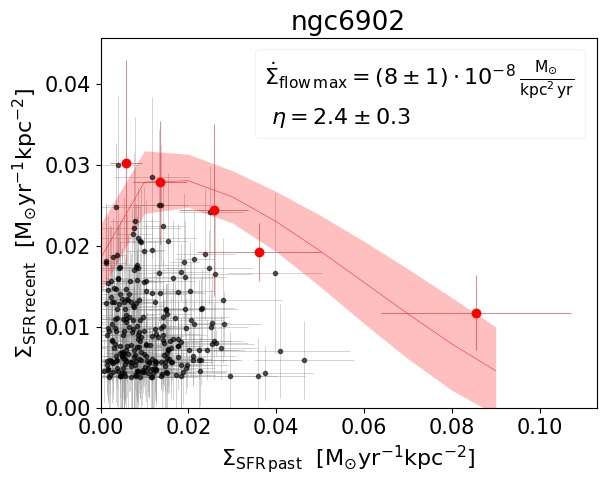}}  
 \subfloat{ \includegraphics[width=0.3\linewidth]{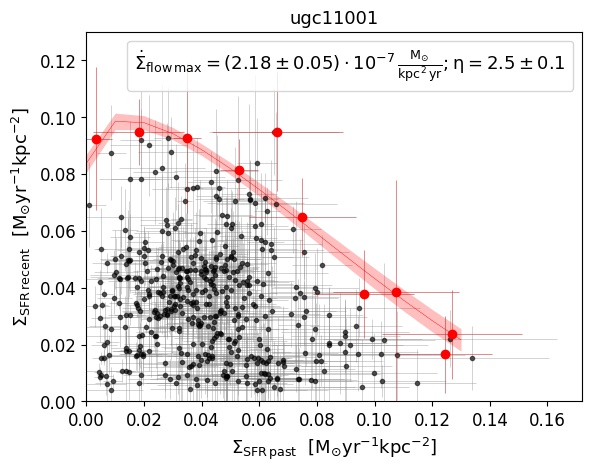}}  \\ 
\subfloat{ \includegraphics[width=0.3\linewidth]{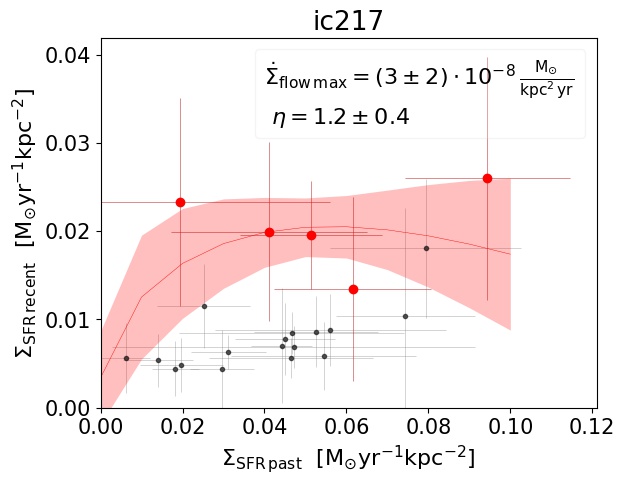}}  
 \subfloat{ \includegraphics[width=0.3\linewidth]{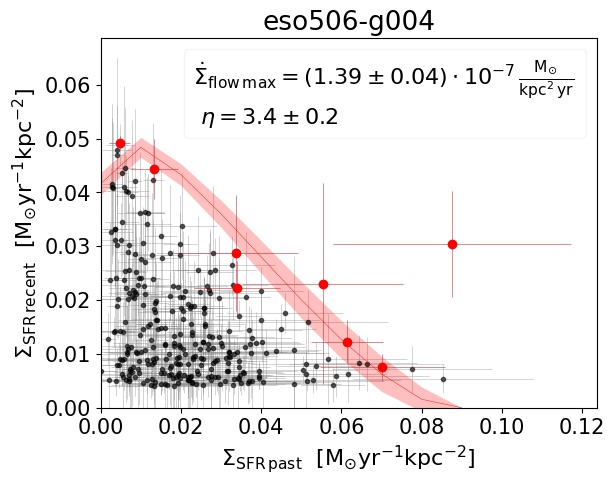}}  
 \subfloat{ \includegraphics[width=0.3\linewidth]{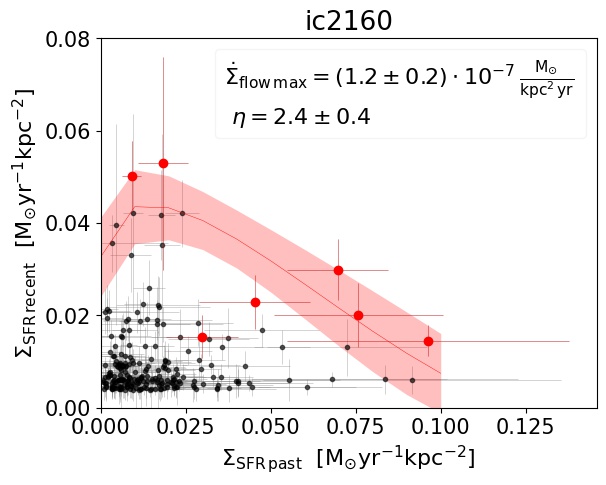}}  \\ 
\subfloat{ \includegraphics[width=0.3\linewidth]{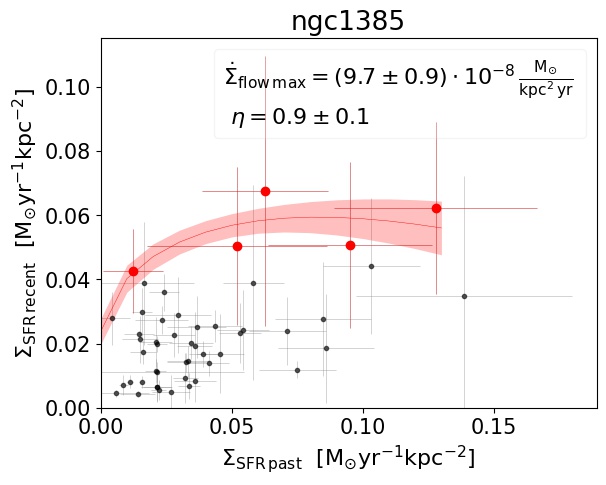}}  
 \subfloat{ \includegraphics[width=0.3\linewidth]{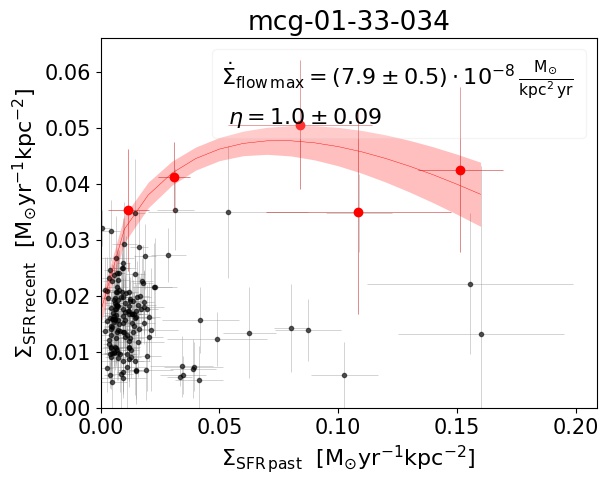}}  
 \subfloat{ \includegraphics[width=0.3\linewidth]{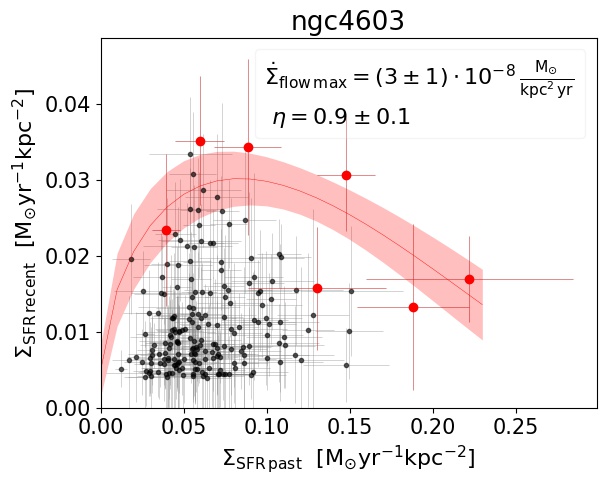}}  \\ 
\contcaption{} 
\end{figure} 
\begin{figure} 
\subfloat{ \includegraphics[width=0.3\linewidth]{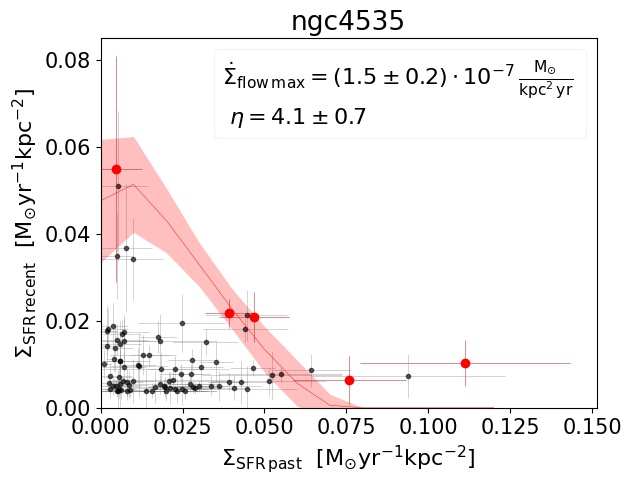}}  
 \subfloat{ \includegraphics[width=0.3\linewidth]{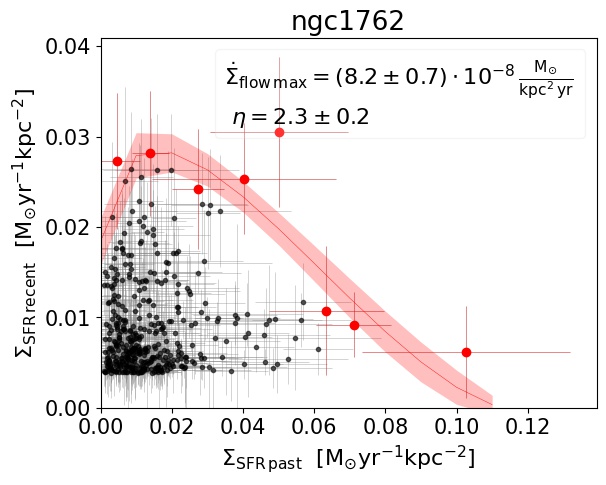}}  
 \subfloat{ \includegraphics[width=0.3\linewidth]{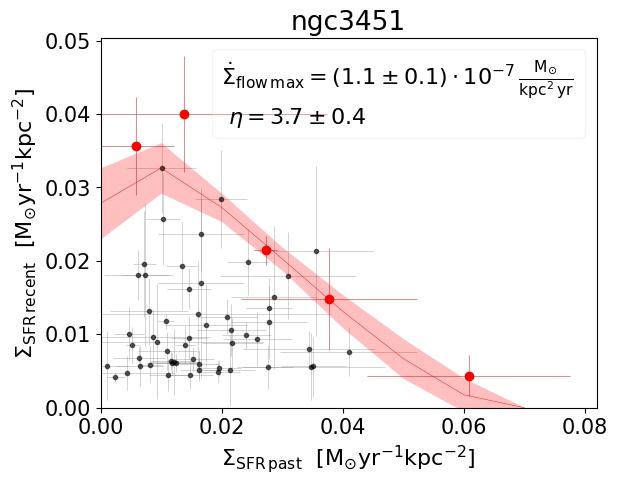}}  \\ 
\subfloat{ \includegraphics[width=0.3\linewidth]{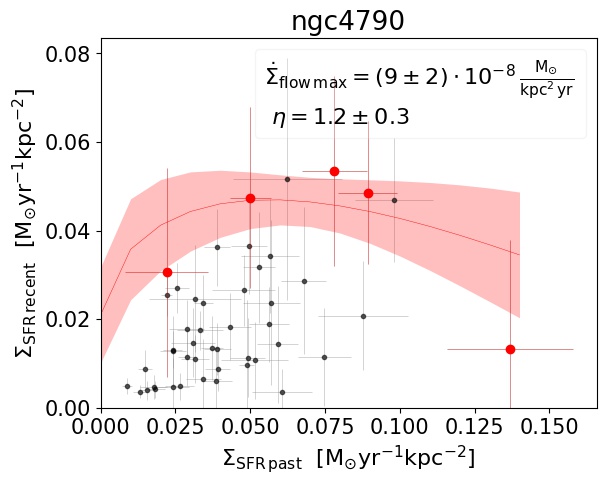}}  
 \subfloat{ \includegraphics[width=0.3\linewidth]{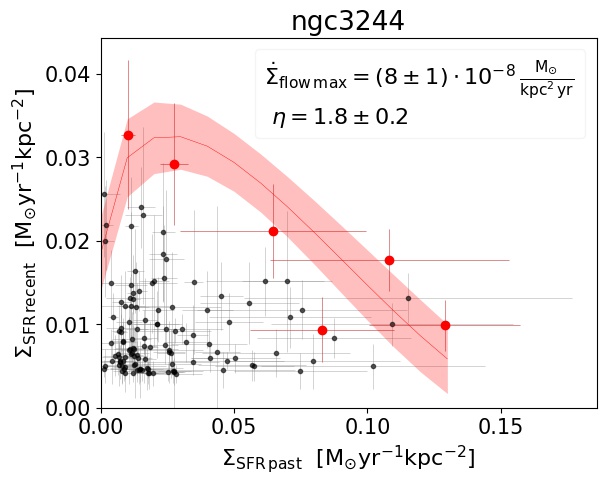}}  
 \subfloat{ \includegraphics[width=0.3\linewidth]{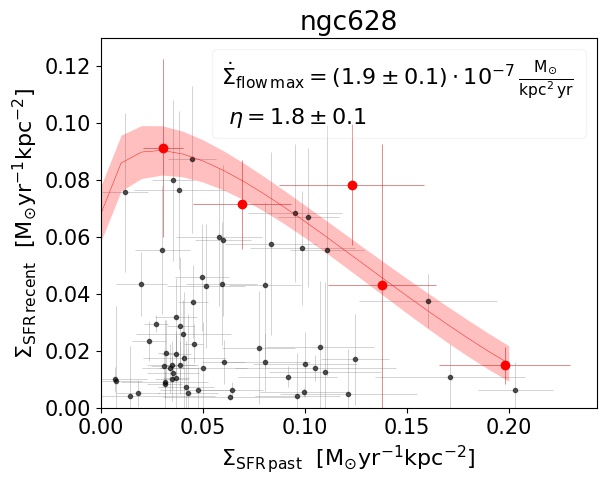}}  \\ 
\subfloat{ \includegraphics[width=0.3\linewidth]{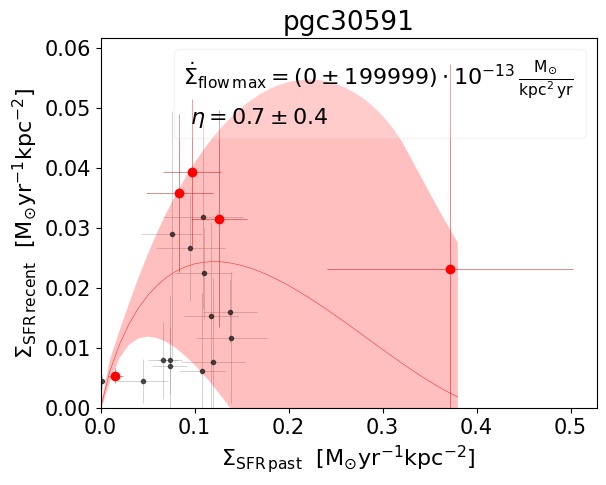}}  
 \subfloat{ \includegraphics[width=0.3\linewidth]{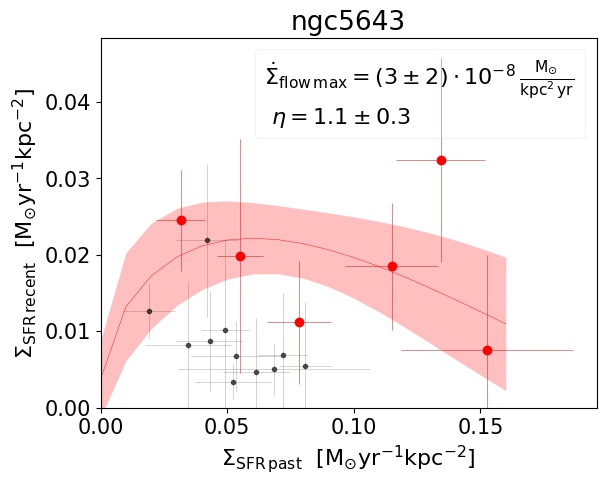}}  
 \subfloat{ \includegraphics[width=0.3\linewidth]{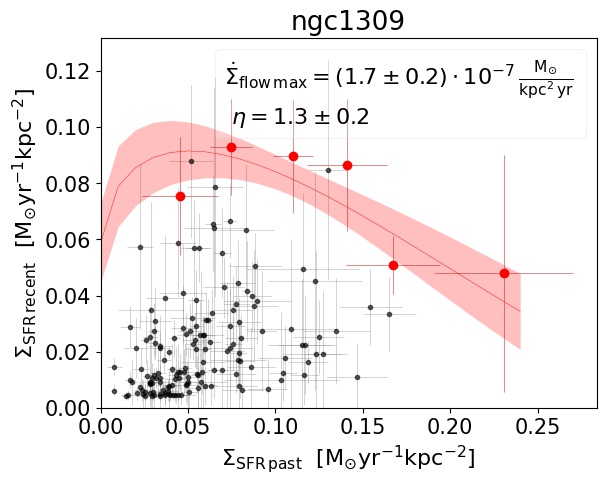}}  \\ 
\subfloat{ \includegraphics[width=0.3\linewidth]{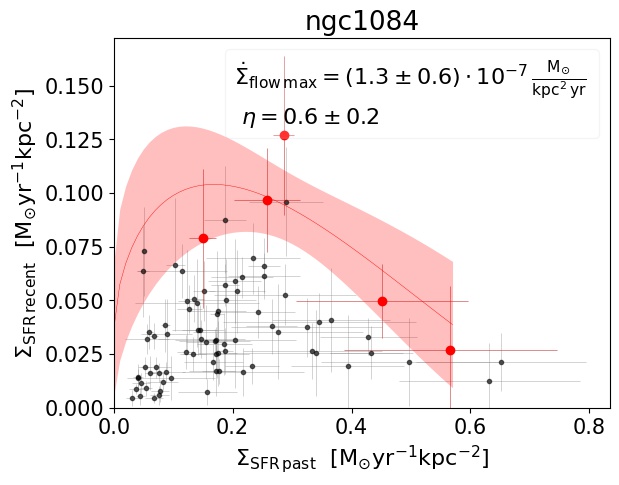}}  
 \subfloat{ \includegraphics[width=0.3\linewidth]{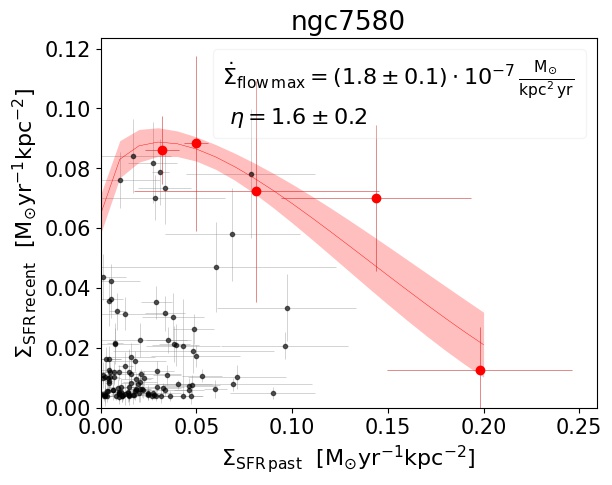}}  
 \subfloat{ \includegraphics[width=0.3\linewidth]{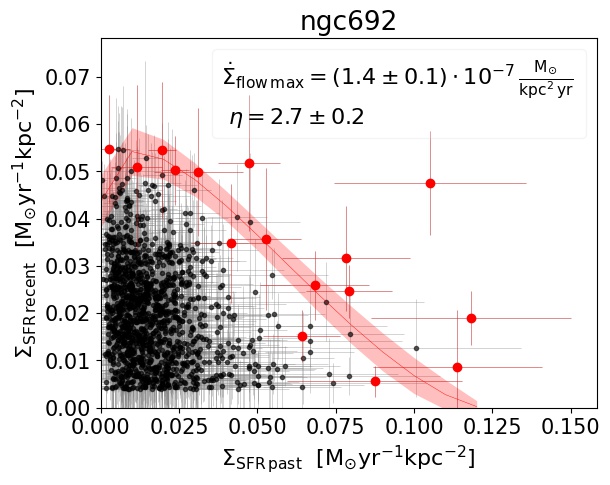}}  \\ 
\contcaption{} 
\end{figure} 
\begin{figure} 
\subfloat{ \includegraphics[width=0.3\linewidth]{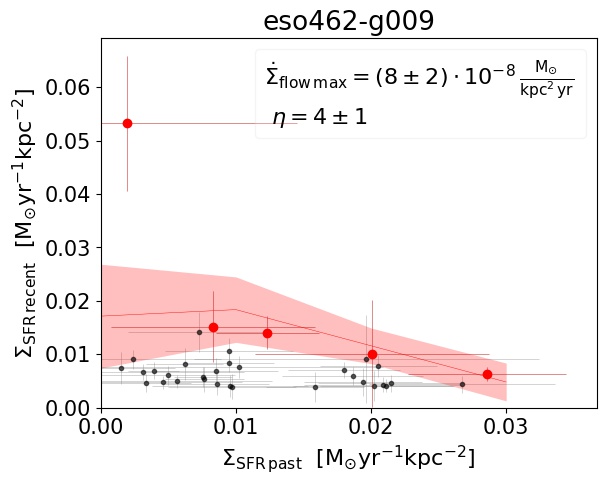}}  
 \subfloat{ \includegraphics[width=0.3\linewidth]{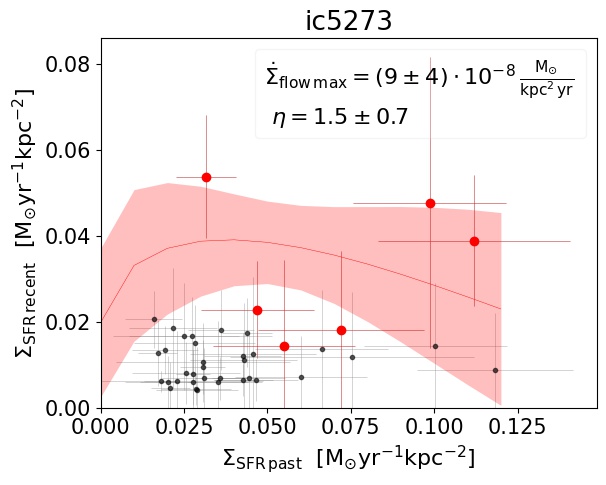}}  
 \subfloat{ \includegraphics[width=0.3\linewidth]{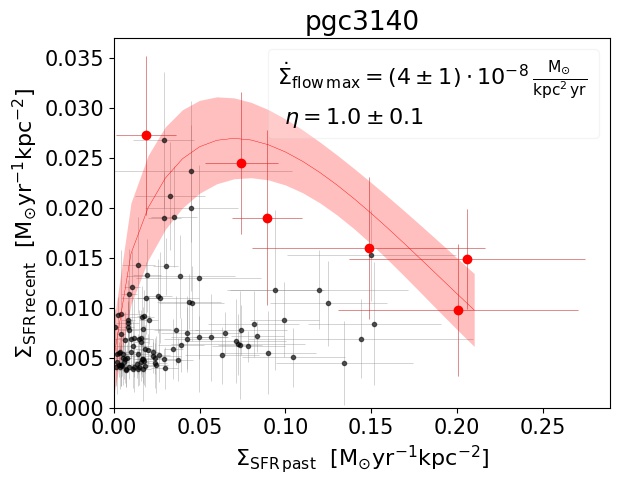}}  \\ 
\subfloat{ \includegraphics[width=0.3\linewidth]{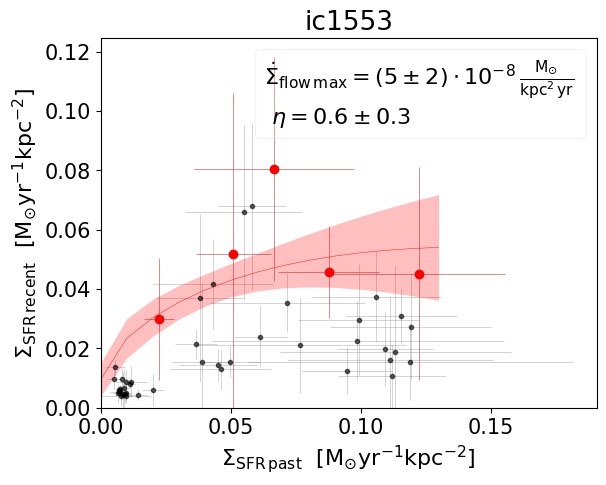}}  
 \subfloat{ \includegraphics[width=0.3\linewidth]{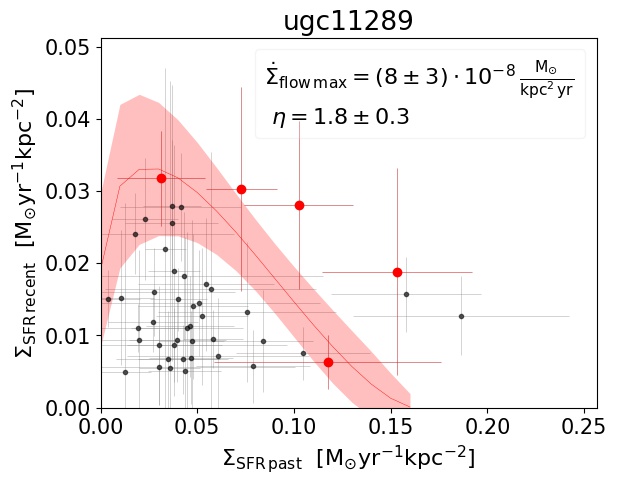}}  
 \subfloat{ \includegraphics[width=0.3\linewidth]{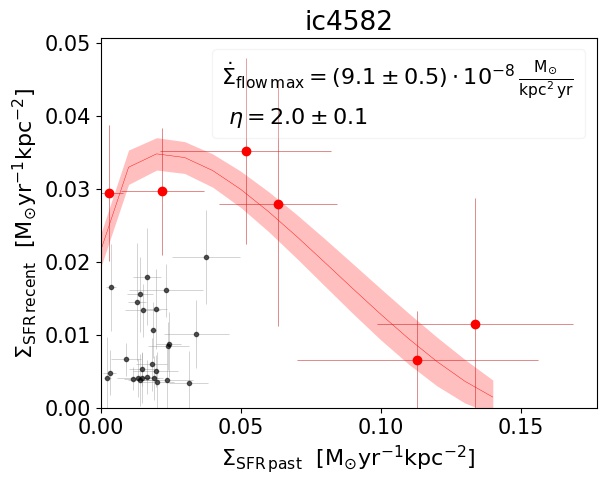}}  \\ 
\subfloat{ \includegraphics[width=0.3\linewidth]{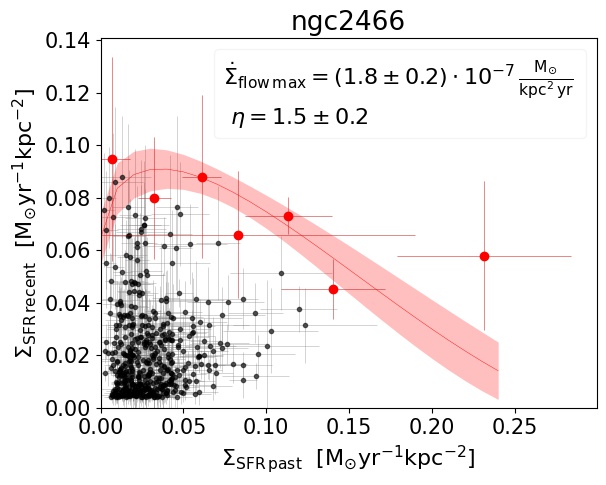}}  
 \subfloat{ \includegraphics[width=0.3\linewidth]{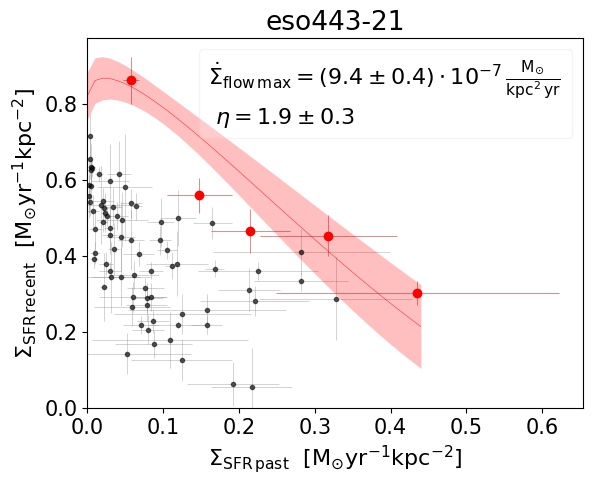}}  
 \subfloat{ \includegraphics[width=0.3\linewidth]{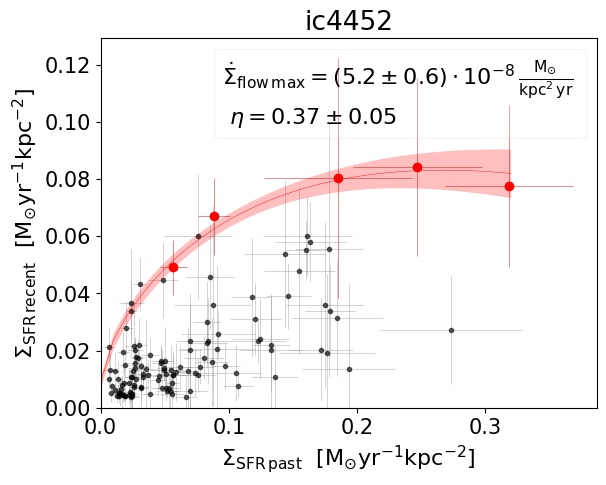}}  \\ 
\subfloat{ \includegraphics[width=0.3\linewidth]{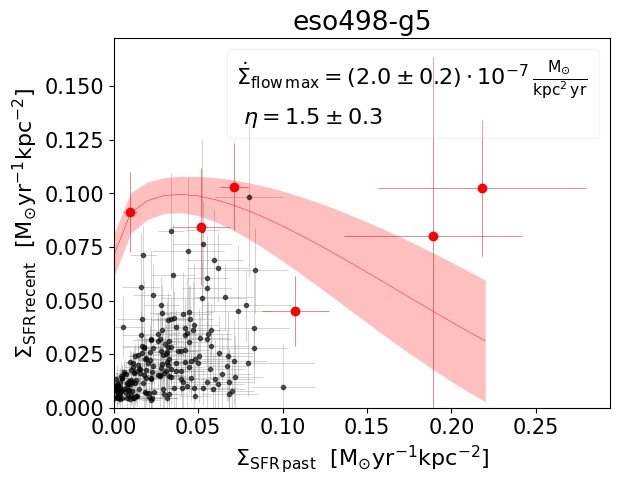}}  
 \subfloat{ \includegraphics[width=0.3\linewidth]{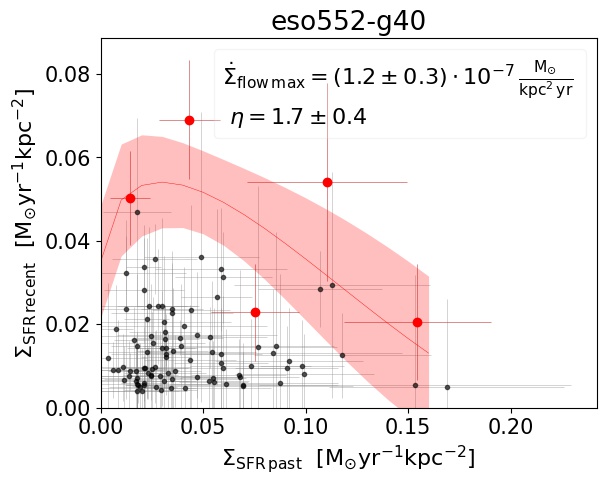}}  
 \subfloat{ \includegraphics[width=0.3\linewidth]{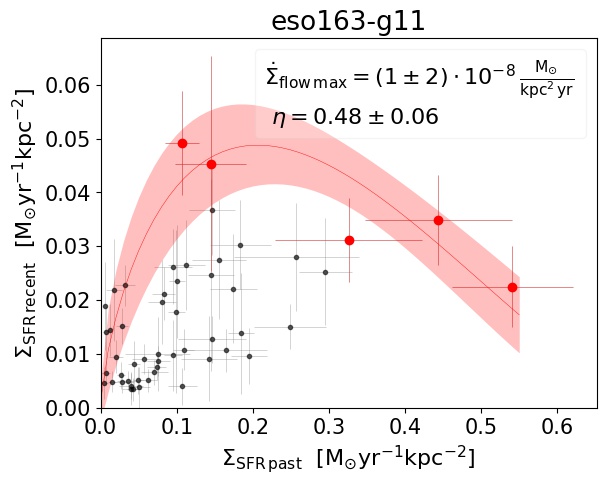}}  \\ 
\contcaption{} 
\end{figure} 
\begin{figure} 
\subfloat{ \includegraphics[width=0.3\linewidth]{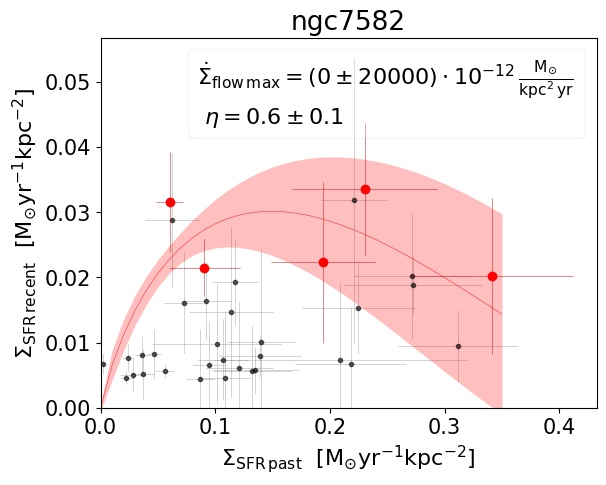}}  
 \subfloat{ \includegraphics[width=0.3\linewidth]{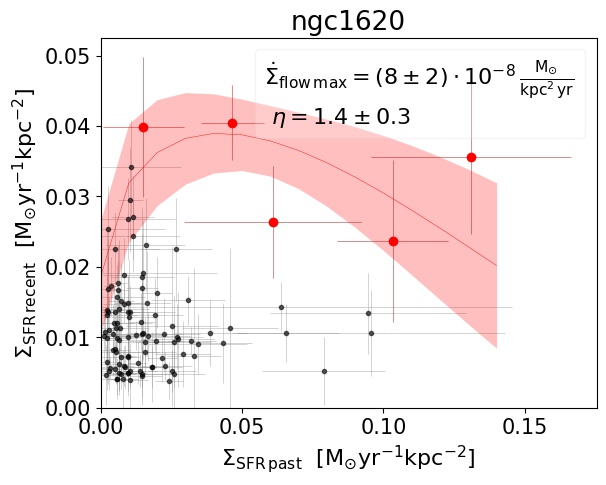}}  
 \subfloat{ \includegraphics[width=0.3\linewidth]{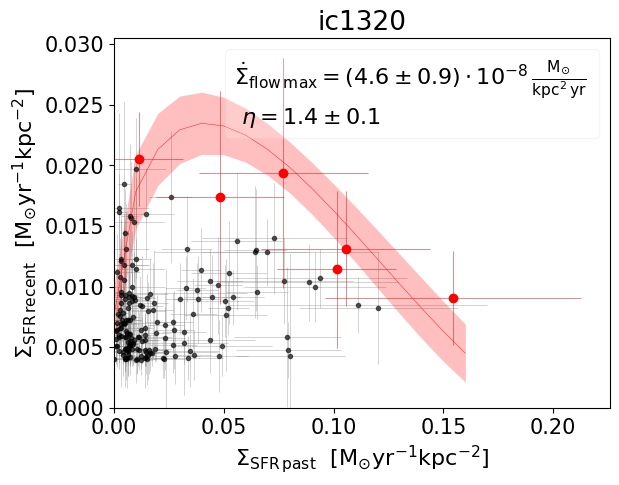}}  \\ 
\subfloat{ \includegraphics[width=0.3\linewidth]{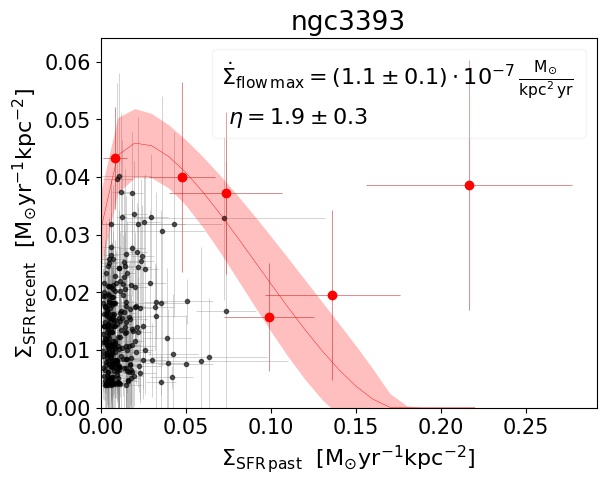}}  
 \subfloat{ \includegraphics[width=0.3\linewidth]{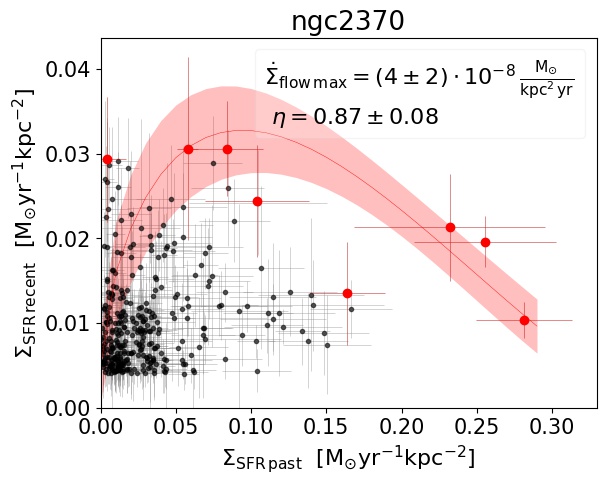}}  
 \subfloat{ \includegraphics[width=0.3\linewidth]{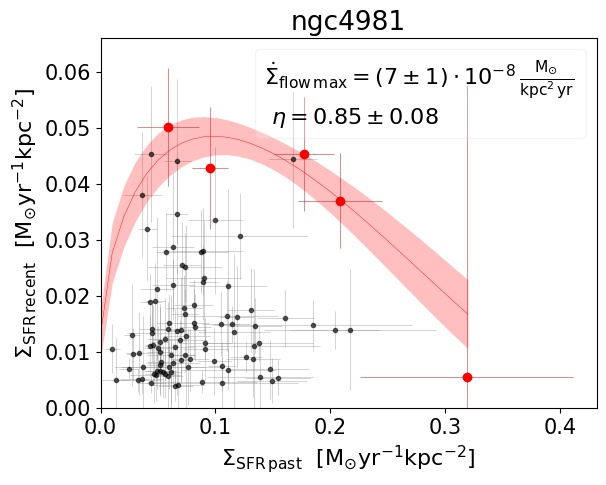}}  \\ 
\subfloat{ \includegraphics[width=0.3\linewidth]{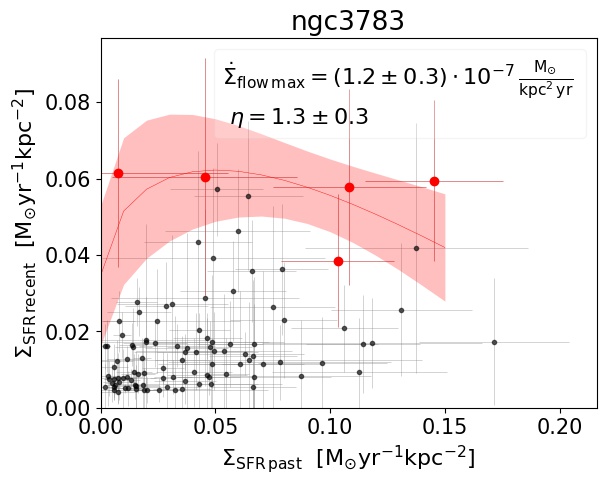}}  
 \subfloat{ \includegraphics[width=0.3\linewidth]{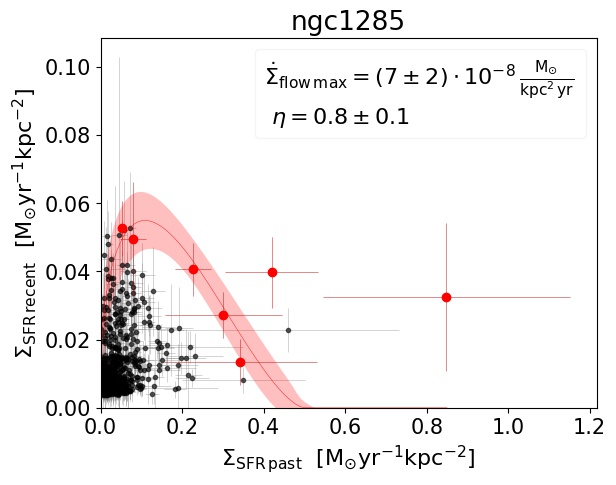}}  
 \subfloat{ \includegraphics[width=0.3\linewidth]{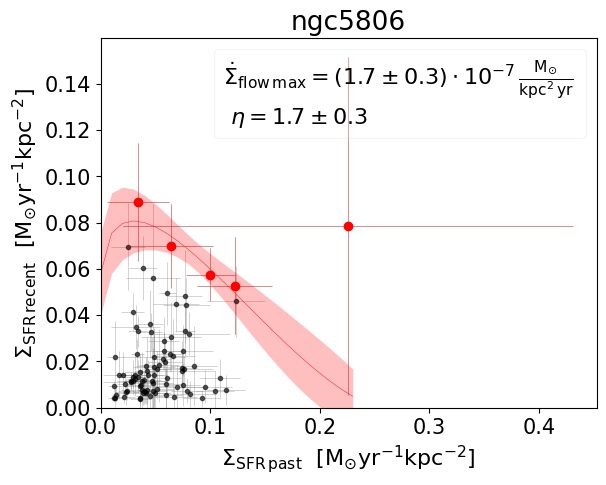}}  \\ 
\subfloat{ \includegraphics[width=0.3\linewidth]{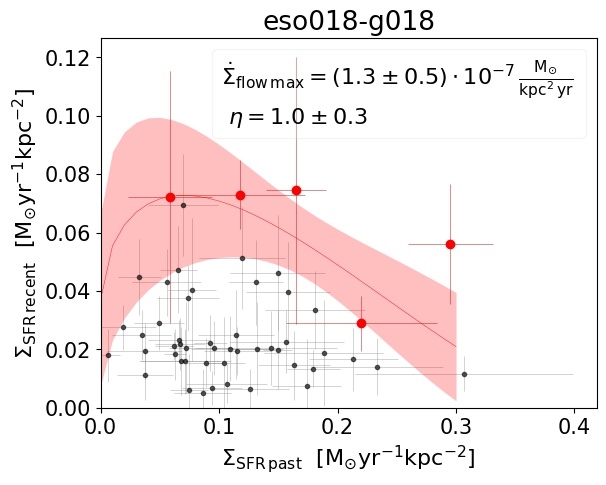}}  
 \subfloat{ \includegraphics[width=0.3\linewidth]{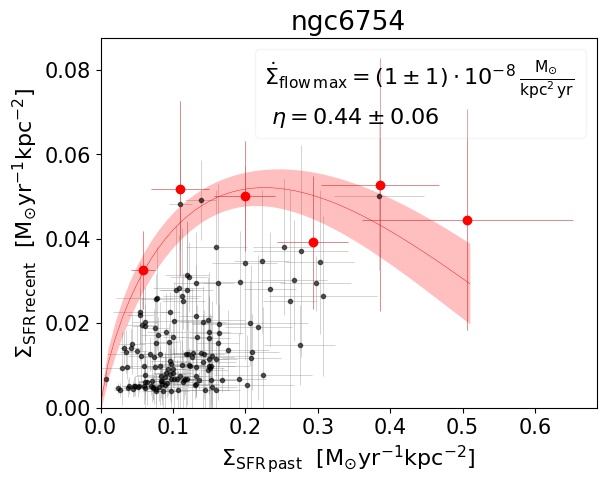}}  
 \subfloat{ \includegraphics[width=0.3\linewidth]{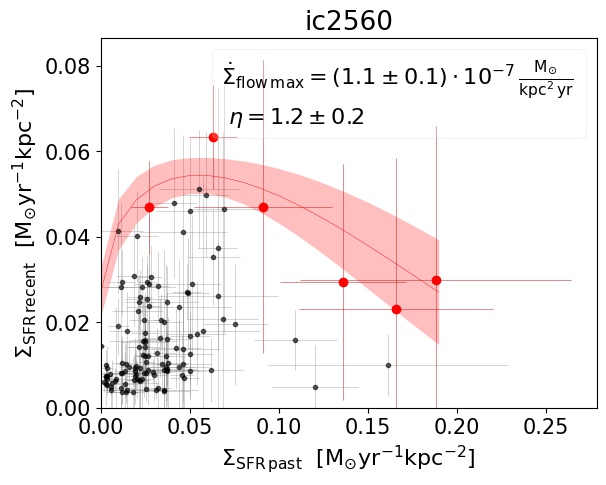}}  \\ 
\contcaption{} 
\end{figure} 
\begin{figure} 
\subfloat{ \includegraphics[width=0.3\linewidth]{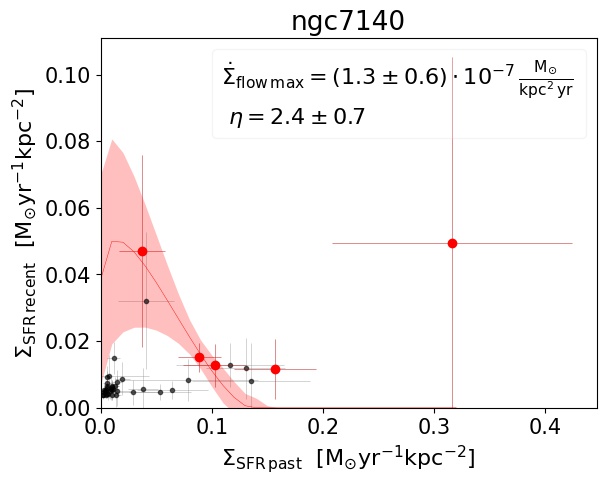}}  
 \subfloat{ \includegraphics[width=0.3\linewidth]{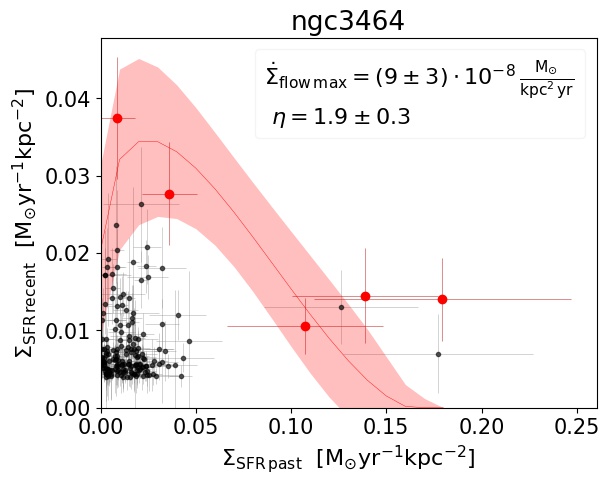}}  
 \subfloat{ \includegraphics[width=0.3\linewidth]{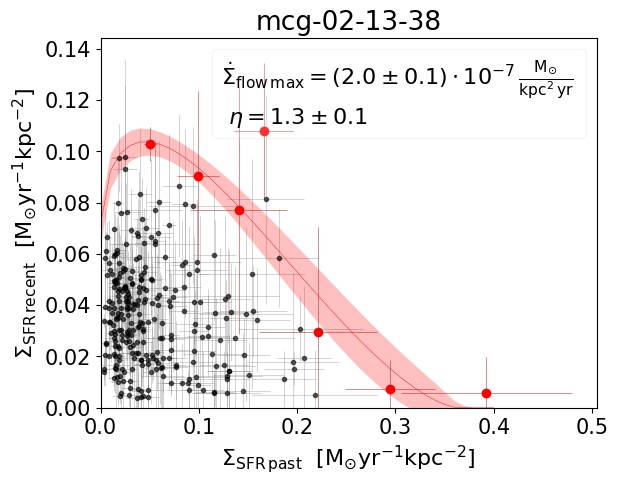}}  \\ 
\subfloat{ \includegraphics[width=0.3\linewidth]{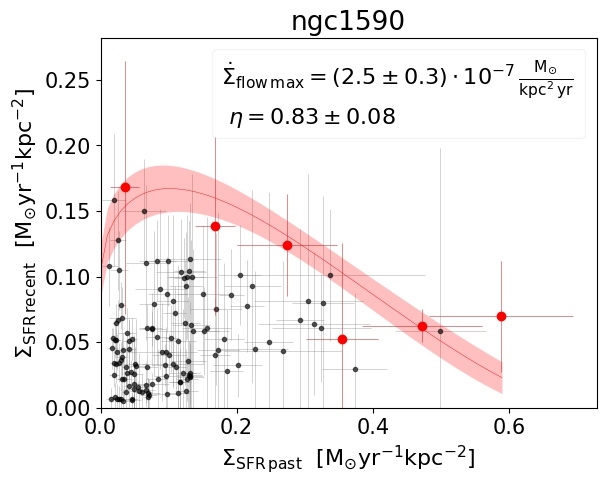}}  
 \subfloat{ \includegraphics[width=0.3\linewidth]{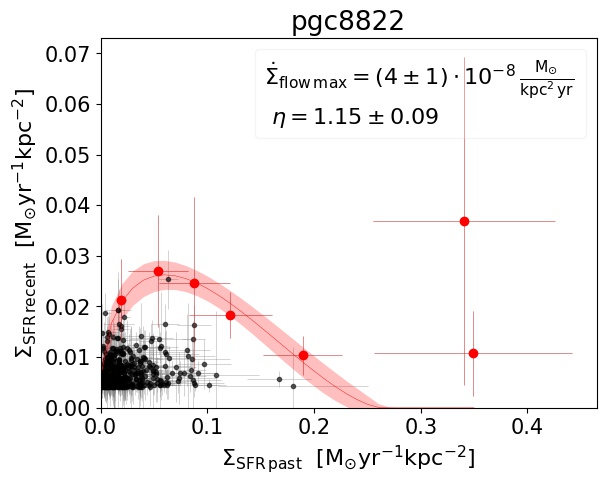}}  
 \subfloat{ \includegraphics[width=0.3\linewidth]{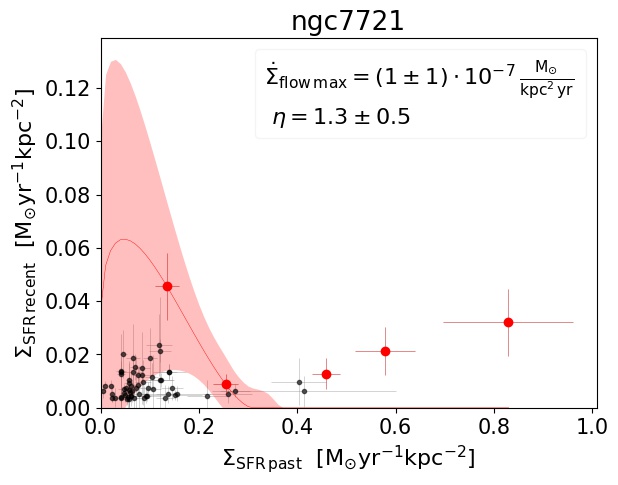}}  \\ 
\subfloat{ \includegraphics[width=0.3\linewidth]{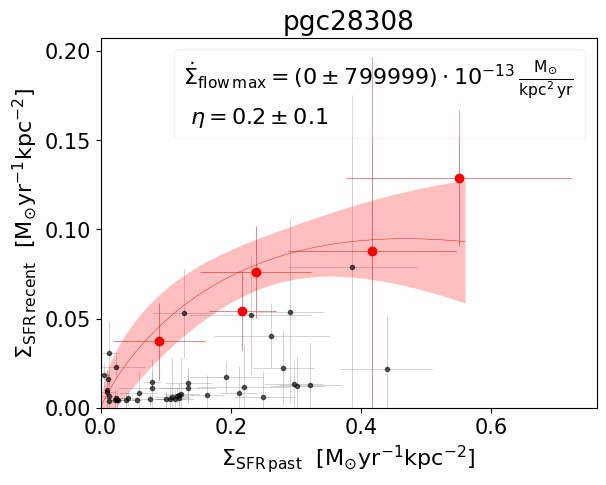}}  
 \subfloat{ \includegraphics[width=0.3\linewidth]{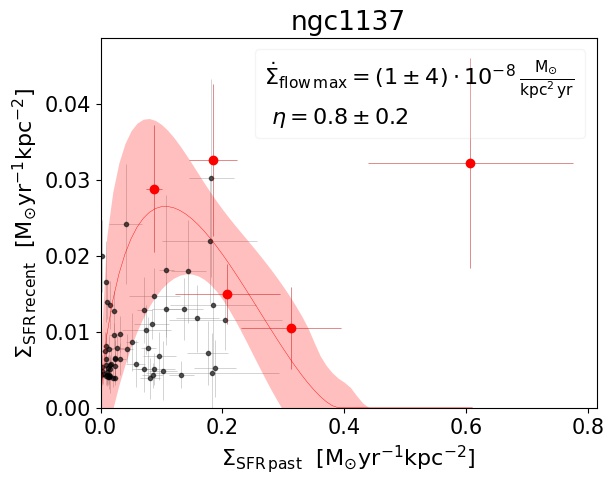}}  
 \subfloat{ \includegraphics[width=0.3\linewidth]{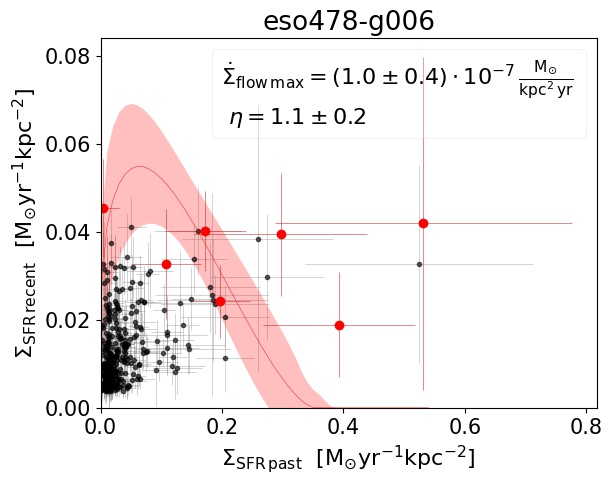}}  \\ 
\subfloat{ \includegraphics[width=0.3\linewidth]{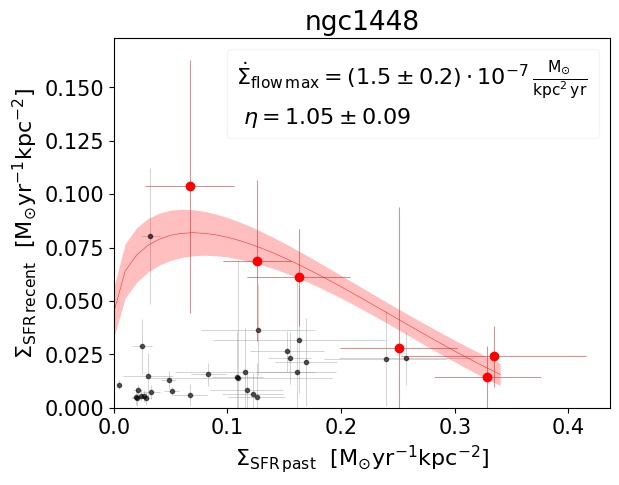}}  
 \subfloat{ \includegraphics[width=0.3\linewidth]{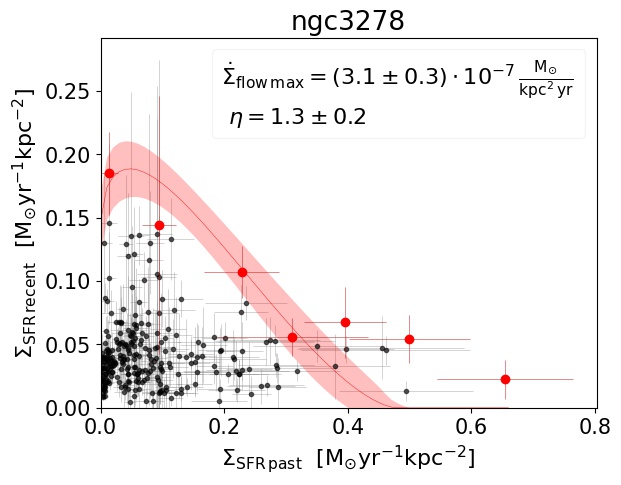}}  
 \subfloat{ \includegraphics[width=0.3\linewidth]{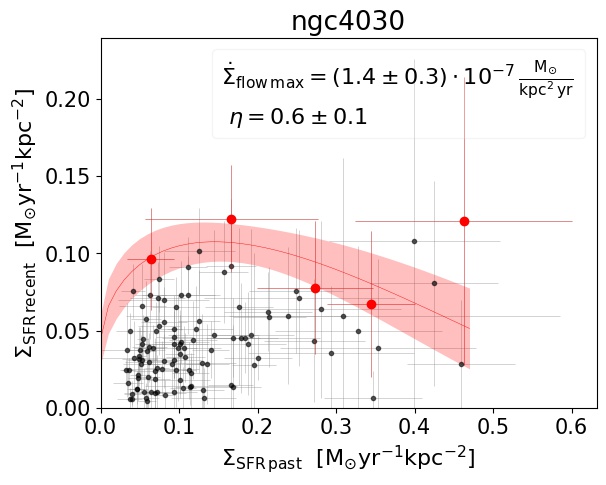}}  \\ 
\contcaption{} 
\end{figure} 
\begin{figure} 
\subfloat{ \includegraphics[width=0.3\linewidth]{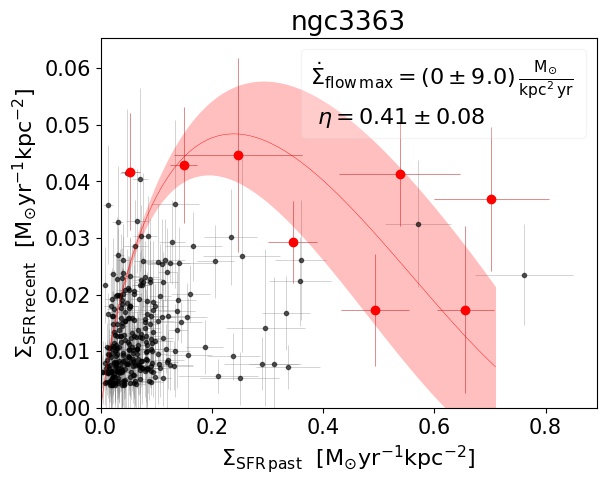}}  
 \subfloat{ \includegraphics[width=0.3\linewidth]{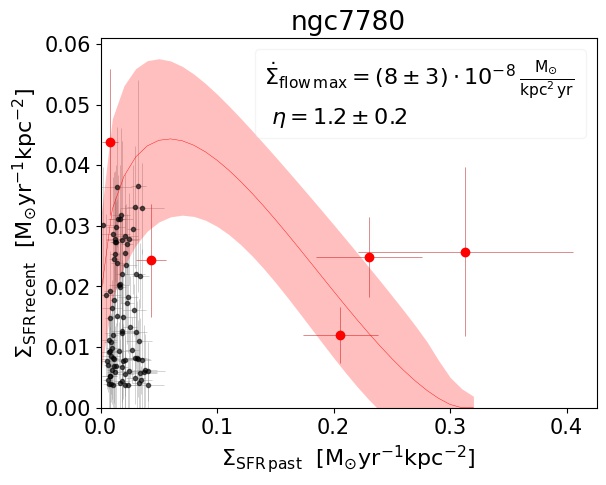}}  
 \subfloat{ \includegraphics[width=0.3\linewidth]{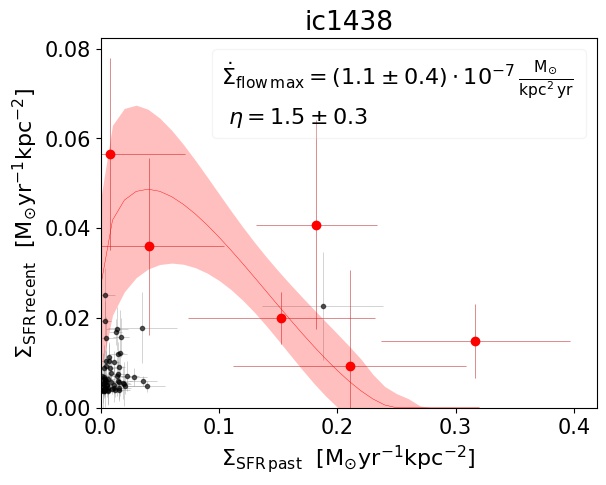}}  \\ 
\subfloat{ \includegraphics[width=0.3\linewidth]{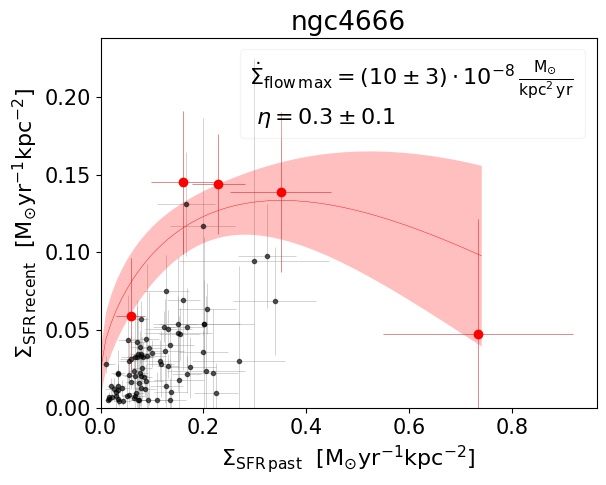}}  
 \subfloat{ \includegraphics[width=0.3\linewidth]{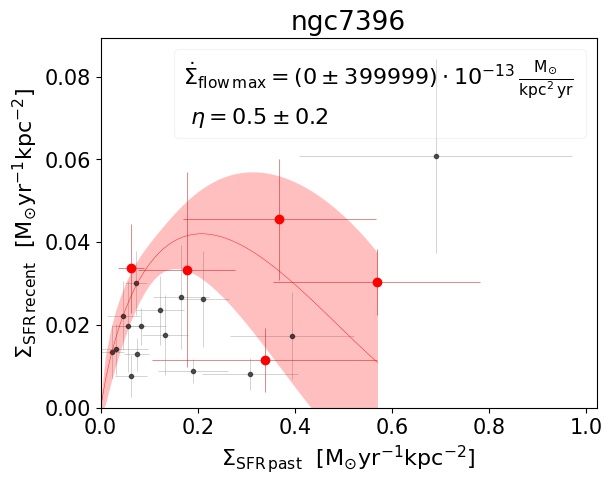}}  
 \subfloat{ \includegraphics[width=0.3\linewidth]{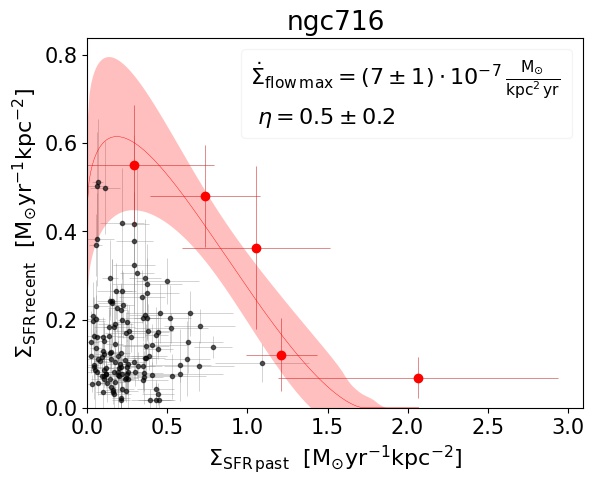}}  \\ 
\contcaption{} 
\end{figure}

% \clearpage
\twocolumn

%%%%%%%%%%%%%%%%%%%%%%%%%%%%%%%%%%%%%%%%%%%%%%%%%%

% Don't change these lines
\bsp	% typesetting comment
\label{lastpage}
\end{document}